\documentclass[a4paper,12pt]{article}
 \pdfoutput=1

\usepackage[pdftex]{graphicx}
\usepackage{subfigure}
\usepackage{float}

\usepackage{xcolor}
\usepackage{multicol}
\usepackage{caption}
\newcommand{\be}{\begin{equation}}
\newcommand{\ee}{\end{equation}}

\newcommand{\bea}{\begin{eqnarray}}
\newcommand{\eea}{\end{eqnarray}}

\newcommand{\p}{\partial}

\newcommand{\nn}{\nonumber \\}
\newcommand{\f}{\frac}
\newcommand{\w}{\wedge}
\newcommand{\ra}{\rightarrow}
\begin{document}
	
\thispagestyle{empty}

\begin{flushright}
%{\bf arXiv:yymm.nnnnn}
\end{flushright}
\begin{center} \noindent \Large \bf 
 %A new black hole solution and 
 Holographic universal relations among transport coefficients
\end{center}

\bigskip\bigskip\bigskip
\vskip 0.5cm
\begin{center}
{ \normalsize \bf   Shesansu Sekhar Pal}

\vskip 0.5 cm

 Department of Physics, Utkal University,  Bhubaneswar, 751004, India\\

\vskip 0.5 cm
\sf { shesansu${\frame{\shortstack{AT}}}$gmail.com }
\end{center}

\centerline{\bf \small Abstract}
In this paper,  we show universal relations among the transport coefficients by calculating the electrical conductivity, thermal conductivity and thermo-electric conductivity in the presence of a chemical potential and magnetic fields for Einstein-Maxwell-dilaton-axion system in arbitrary but even dimensional bulk spacetime as well as for Einstein-DBI-dilaton-axion system in $3+1$ dimensional bulk spacetime. Moreover, we have also obtained a  new hyperscale violating black hole solution with finite charge density and magnetic fields but with a trivial dilaton field at IR. 
\newpage

\section{Introduction}

It is known from the studies of AdS/CFT correspondence that the charged black holes in gravitational theory corresponds to states at finite temperature with nonzero charge density or non-zero chemical potential in the dual  field theory \cite{Maldacena:1997re}. The black hole is made charged via the Maxwell field which is dual to the current of a global U(1) symmetry in the dual field theory. The AdS/CFT correspondence has been used to understand the transport coefficients
of holographic matter at finite temperature with a finite density and magnetic field in \cite{Hartnoll:2007ih, Hartnoll:2007ip}. Studies of holographic matter is reviewed, recently, in \cite{Hartnoll:2016apf}. It is suggested in \cite{Iqbal:2008by} that in the low frequency limit, the transport coefficients can be calculated by evaluating some geometric quantities at the horizon.

The longitudinal electrical conductivity of Einstein-Maxwell-dilaton-axion system with explicit break down of the spatial translational symmetry   is computed for AdS spacetime in  \cite{Andrade:2013gsa}.  Based on the result of computation, it is suggested in \cite{Liu:2020rrn} to write it as 
\be
\sigma_L=(\mu L)^{d-3}w_0 ~\left(g\bigg( \f{T_H}{\mu}, \f{k}{\mu} \bigg)\right)^{d-3
}\left[1+\f{(d-2)^2w_0\mu^2}{\psi_0k^2L^2}\right], 
\ee
A prescription is given   to calculate the transport coefficients with momentum dissipation for holographic matter at the horizon in \cite{Donos:2014cya} and it matches with the computation made in \cite{Andrade:2013gsa}. Further studies are made in \cite{Blake:2017qgd, Blake:2014yla}. 
We have  calculated the longitudinal thermo-electric and thermal conductivity and the result reads as
\bea
\alpha_L&=&(\mu L)^{d-1} ~\left(g\bigg( \f{T_H}{\mu}, \f{k}{\mu} \bigg)\right)^{d-2} \left[\f{4\pi(d-2)w_0}{\psi_0k^2 L^2}\right],\nn
\kappa_L&=&(\mu L)^{d-1} ~\left(g\bigg( \f{T_H}{\mu}, \f{k}{\mu} \bigg)\right)^{d-1}\left(\f{16\pi^2 T_H  }{\psi_0k^2}\right)\left[1+\f{(d-2)^2w_0\mu^2}{\psi_0k^2L^2}\right]^{-1}
\eea
where the form of the function $g(x,~y)$ is to be determined by solving it

\be
\mu L~~ g\bigg( \f{T_H}{\mu}, \f{k}{\mu}  \bigg)=\f{r_h}{L}
\ee
in terms of the size of the horizon, $r_h$, and the horizon size has to be determined by solving   eq(\ref{temp_b_charge_density_dissipation}) by setting the  magnetic field to zero.  It just follows that generically it is not possible to separate the incoherent production of the particle-hole pairs from the momentum dissipation due to lattice effects. 

\paragraph{AdS spacetime:} In this paper, we have revisited  the charged AdS black holes solutions with planar horizon in arbitrary spacetime dimensions in Einstein-Maxwell-dilaton-axion system  and found that the   celebrated Wiedemann-Franz relation, unfortunately, does not hold. However, there exists a relation among the transport coefficients. The relation involves electrical conductivity, thermo-electric conductivity,  thermal conductivity, chemical potential  and the temperature, which  reads as

\be
\f{\mu^2}{k^2}\f{\kappa_L\sigma_L}{T_H \alpha^2_L}=\f{\psi_0 L^2}{(d-2)^2 w_0}={\rm constant}
\ee

This relation involves quantities which are defined at the horizon such as transport coefficients and the chemical potential at the boundary (the behavior of the gauge potential)  means the above product is not necessarily universal \cite{Davison:2016ngz}. This is true as it
works only for the AdS spacetime.

\paragraph{Universal relations:} As soon as we turn on the magnetic field the transport coefficients takes the form as written in
eq(\ref{transport_b}) and eq(\ref{transport_kappa}). It is easy to notice that the non-vanishing behavior of longitudinal electrical conductivity, thermo-electric and thermal conductivities are governed directly by chemical potential, momentum dissipation  and temperature,  respectively. However,  the non-vanishing behavior of transverse conductivities are governed by temperature, chemical potential and  magnetic field.

In fact, it follows upon closer inspection that the transport coefficients evaluated at the horizon are not all independent. There exists an interesting  relation among the transport coefficients

\be\label{holo_transport_relation}
\fcolorbox{lightgray}{white}{
$\displaystyle
   T_H\f{\alpha_{11}(r_h)}{\overline\kappa_{12}(r_h)}\f{\alpha_{12}(r_h)}{\sigma_{11}(r_h)}=1$
},
\ee
which in turn gives a relation that involves all the transport coefficients except the longitudinal thermal conductivity at the horizon 
\be
\fcolorbox{teal}{white}{
$\displaystyle
  T_H\left[ \sigma_{11}\sigma_{12}(\alpha^2_{11}-\alpha^2_{12})-\alpha_{11}\alpha_{12}(\sigma^2_{11}-\sigma^2_{12})\right]=\sigma_{11}\kappa_{12}(\sigma^2_{11}+\sigma^2_{12})$}.
\ee
It says the off-diagonal component of the thermal conductivity matrix, which is the transverse thermal conductivity,  can be determined completely in terms of  the electrical and thermo-electric conductivities. It holds irrespective of the precise detail of the black hole spacetime.  We have checked that the relation eq(\ref{holo_transport_relation}) also holds for the Einstein-DBI-dilaton-axion system studied in \cite{Pal:2019bfw, Pal:2020gsq}.

We also show  relations
\bea\label{holo_transport_relation_II}
&&\fcolorbox{teal}{white}{
	$\displaystyle
\f{\sigma_{11}(r_h)}{\alpha_{12}(r_h)}\f{\vartheta_{11}(r_h)}{\rho_{11}(r_h)}=-\f{\rho}{16\pi G B}=T_H\f{\vartheta_{11}(r_h)}{\rho_{11}(r_h)}\f{\alpha_{11}(r_h)}{\overline\kappa_{12}(r_h)}=-\left(\f{Q}{V_{d-1}B}\right)$},\nn
&&\fcolorbox{teal}{white}{
	$\displaystyle T_H\f{\alpha_{12}(r_h)}{\sigma_{11}(r_h)}\f{\vartheta_{12}(r_h)}{\kappa_{12}(r_h)}=\left(\f{16\pi G B}{\rho}\right)=\f{\overline\kappa_{12}(r_h)}{\alpha_{11}(r_h)}\f{\vartheta_{12}(r_h)}{\kappa_{12}(r_h)}=\left(\f{V_{d-1} B}{Q}\right)$},
\eea
where $\vartheta_{11}, \vartheta_{12}, \rho_{ij}$ and $Q$ are the  Seebeck coefficient, Nernst response,   resistivity coefficients and  the electric charge, respectively. It is interesting to note that the relation eq(\ref{holo_transport_relation_II}) follows from eq(\ref{holo_transport_relation}) via the duality relation.

%The sum of the squares of the longitudinal electrical conductivity as well as the transverse electrical conductivity obeys 
%\be\label{circle}
%\sigma^2_{11}(r_h)+\sigma^2_{12}(r_h)=(W(\phi(r_h)) )^2 ~ h^{d-3}(r_h).
%\ee
%It is important to note that this relation works only when the charge density, magnetic field and the geometric quantities are related by $\rho=B~~  W(\phi(r_h))  ~~ h^{\f{d-3}{2}}(r_h) $, which is the self-dual point\footnote{The self-dual point is defined as the point for which the product of the charge density and the magnetic field remain unchanged under the transformation as written in eq(\ref{duality}).}.

\paragraph{New black hole solution:} In order to include the effect of the charge density at finite temperature at IR, it is suggested in \cite{Tarrio:2011de, Alishahiha:2012qu} to include more than one  U(1) gauge fields. 
In this paper, we present 
a new class of solution at non-zero temperature and charge density in the presence of   magnetic field by considering only one gauge field. Moreover, we do not need a non-trivial profile of the dilaton at IR. The solution is characterized by two exponents $z$ and $\gamma$. Non-zero value of  $\gamma$ essentially describes the scale violation.

The paper is organized as follows. In {\bf  section 2}, we shall present the Einstein-Maxwell-dilaton-axion system and calculate the currents after including the geometric and gauge fields fluctuations. In {\bf  section 3}, we shall calculate the transport coefficients.  In {\bf  section 4}, we shall introduce a transformation that interchanges the charge density with the magnetic field and  study its impact on the transport coefficients. We prove  the relation eq(\ref{holo_transport_relation}) and eq(\ref{holo_transport_relation_II}). In {\bf  section 5}, we study the behavior of the transport coefficients versus temperature and the magnetic field for AdS spacetime as well as for the scale violating spacetime. The thermodynamics of the AdS spacetime  is studied in the Appendix A. The relations among the transport coefficients, i.e.,  eq(\ref{holo_transport_relation}) and eq(\ref{holo_transport_relation_II}) are proved  for the planar black holes in the Einstein-DBI-dilaton-axion system  in Appendix B.

\section{The system}
The system that we shall be considering for our study of the
transport coefficients at finite chemical potential and magnetic field
 %computation of the electrical conductivity, thermal conductivity and thermo-electric conductivity in the presence of charged density 
 in arbitrary spacetime dimensions is as follows.

The action that we shall be considering involves metric, dilaton, $(d-1)$ number of  axions, gauge field which takes the following form in $d+1$ spacetime dimensions
\bea\label{action}
S_{bulk}&=&\f{1}{2\kappa^2}\int d^{d+1} x\sqrt{-g}\bigg[ {\cal R}-2\Lambda-\f{1}{2}(\p \phi)^2-V(\phi)-\f{W(\phi)}{4}F_{MN}
F^{MN}\nn&-&\f{\Psi(\phi)}{2}\Bigg((\p \chi_1)^2+(\p \chi_2)^2+\cdots+(\p \chi_{d-1})^2 \Bigg)\Bigg].
\eea
We shall take $d$ to be odd. 
The equation of motion that follows for the metric tensor  are 
\bea
&&{\cal R}_{MN}-\f{\Psi(\phi)}{2}\Bigg(\p_M \chi_1\p_N\chi_1+\p_M \chi_2\p_N\chi_2+\cdots+\p_M \chi_{d-1}\p_N\chi_{d-1}\Bigg) -\f{W(\phi)}{2} {F_M}^L F_{NL}+\nn&&\f{W(\phi)}{4(d-1)}g_{MN}F^{KL}F_{KL}-\f{[V(\phi)+2\Lambda]}{(d-1)}g_{MN}-
\f{1}{2}\p_M\phi\p_N\phi=0.
\eea

 For the gauge field 

\be
\p_{M}\left[\sqrt{-g} W(\phi) F^{MN} \right]=0,\quad {\rm where}\quad F_{MN}=\p_M a_N-\p_N a_M.
\ee

The equation of motion associated to the scalar field is
\bea\label{scalar_eom}
&&\p_{M}\bigg(\sqrt{-g}\p^M\phi \bigg)-\sqrt{-g}\f{dV(\phi)}{d\phi}-\f{\sqrt{-g}}{2}\f{d\Psi(\phi)}{d\phi}\bigg[(\p \chi_1)^2+(\p \chi_2)^2+\cdots+(\p \chi_{d-1})^2\bigg]\nn&&-\f{\sqrt{-g}}{4}\f{dW(\phi)}{d\phi}F_{MN}F^{MN}=0.
\eea

The equation of motion associated to axions, $\chi_i$'s,  are
\be
\p_{M}\bigg(\sqrt{-g}\Psi(\phi)\p^M\chi_i \bigg)=0,\quad {\rm for}~~ i= 1,2, \cdots, (d-1).
\ee

\paragraph {Solution:} To read out the solution of the above equations, let us take the following  ansatz of the metric and other matter fields as follows
\bea\label{solution_gen}
ds^2_{d+1}&=&-g_{tt}(r)dt^2+g_{xx}(r)\left( dx^2_1+\cdots+dx^2_{d-1}\right)+g_{rr}(r)dr^2,\quad \phi=\phi(r),\quad\chi_i=kx_i,\nn
\quad F&=&a'_t(r)dr\w dt+B\left( dx_1\w dx_2+dx_3\w dx_4+\cdots+dx_{d-2}\w dx_{d-1}\right).
\eea
We shall consider the case for which $d-1$ is even, it means $d$ odd. 
In what follows we shall write $g_{tt}(r)=U_1(r)~~g_{xx}(r)=h(r)$ and $g_{rr}(r)=\f{1}{U_2(r)}$. After a bit of calculations  with such a choice of the geometry allows us to write   the Ricci tensor as  follows 

\bea
{\cal R}_{tt}&=&\f{(d-1)U_2h'U'_1}{4h}+\f{U_2U''_1}{2}-\f{U_2U'^2_1}{4U_1}+\f{U'_1U'_2}{4},\nn
{\cal R}_{rr}&=&\f{(d-1)h'^2}{4h^2}+\f{U'^2_1}{4U^2_1}-
\f{(d-1)h'U'_2}{4hU_2}-\f{U'_1U'_2}{4U_1U_2}-\f{(d-1)h''}{2h}-\f{U''_1}{2U_1}
 ,\nn
 {\cal R}_{ij}&=&-\delta_{ij}\left[\f{(d-3)U_2h'^2}{4h}+\f{U_2h'U'_1}{4U_1}+\f{U_2h''}{2}+\f{h' U'_2}{4} \right].
\eea

\paragraph{Equation of motion:} The equation of motion associated to geometry are
\bea
&&{\cal R}_{tt}-\f{W(d-2)}{2(d-1)}U_2a_t^{'2}+\f{(V+2\Lambda)}{(d-1)}U_1-\f{WU_1B^2}{4h^2}=0,\nn
&&{\cal R}_{ij}-\f{\Psi}{2}k^2\delta_{ij}-\f{WU_2}{2(d-1)U_1}ha_t^{'2}\delta_{ij}-\f{(V+2\Lambda)}{(d-1)}h\delta_{ij}-\f{WB^2}{4h}\delta_{ij}=0,\nn
&&{\cal R}_{rr}+\f{(d-2)}{2(d-1)}\f{W}{U_1}a_t^{'2}-
\f{(V+2\Lambda)}{(d-1)U_2}-\f{1}{2}\phi'^2+\f{WB^2}{4h^2U_2}=0,
\eea

writing it out explicitly gives
\bea\label{eom}
&&\f{(d-1)U_2h'U'_1}{4h}+\f{U_2U''_1}{2}-\f{U_2U'^2_1}{4U_1}+\f{U'_1U'_2}{4}-\f{W(d-2)}{2(d-1)}U_2a_t^{'2}+\f{(V+2\Lambda)}{(d-1)}U_1-\f{WU_1B^2}{4h^2}=0,\nn
&&\f{(d-3)U_2h'^2}{4h}+\f{U_2h'U'_1}{4U_1}+\f{U_2h''}{2}+\f{h' U'_2}{4} +\f{\Psi}{2}k^2+\f{WU_2}{2(d-1)U_1}ha_t^{'2}+\f{(V+2\Lambda)}{(d-1)}h+\f{WB^2}{4h}=0,\nn
&&\f{(d-1)h'^2}{4h^2}+\f{U'^2_1}{4U^2_1}-
\f{(d-1)h'U'_2}{4hU_2}-\f{U'_1U'_2}{4U_1U_2}-\f{(d-1)h''}{2h}-\f{U''_1}{2U_1}+\nn&&\f{(d-2)}{2(d-1)}\f{W}{U_1}a_t^{'2}-
\f{(V+2\Lambda)}{(d-1)U_2}-\f{1}{2}\phi'^2+\f{WB^2}{4h^2U_2}=0.
\eea

These equations yield a constraint equation
\bea\label{constraint}
&&2h^2\Bigg(2U_1(V+2\Lambda)+U_2(Wa'^2_t-U_1\phi'^2)\Bigg)\nn&=&-(d-1)\left[ 2h(k^2U_1\psi+U_2h'U'_1)+U_1(B^2W+(d-2)U_2h'^2)\right]
\eea

The equation of motion for the gauge potential can be solved and it gives
\be
a'_t(r)=\rho\f{\sqrt{U_1(r)}}{\sqrt{U_2(r)}W(\phi(r))h^{\f{d-1}{2}}(r)},
\ee
where $\rho$ is a constant and which upon integration gives 
%\be
%A_t(r)=A_t(r_h)+\rho\int^r_{r_h}\f{\sqrt{U_1(r)}}{\sqrt{U_2(r)}W(\phi(r))h^{\f{d-1}{2}}(r)}.
%\ee
 
\be
\mu=A_t(r=\infty)=\rho\int^{\infty}_{r_h}dr \f{\sqrt{U_1(r)}}{\sqrt{U_2(r)}W(\phi(r))h^{\f{d-1}{2}}(r)}.
\ee
This allows the charge susceptibility as \cite{Iqbal:2008by}
\be
\chi_e\equiv\f{\rho}{\mu}=\left( \int^{\infty}_{r_h} dr \f{\sqrt{U_1(r)}}{\sqrt{U_2(r)}W(\phi(r))h^{\f{d-1}{2}}(r)}\right)^{-1}.
\ee

The equation of motion of the dilaton field is
\bea\label{scalar_eom}
&&\p_{r}\bigg(\sqrt{U_1U_2}h^{\f{d-1}{2}}\phi' \bigg)-\f{\sqrt{U_1}}{\sqrt{U_2}}h^{\f{d-1}{2}}\bigg[\f{dV(\phi)}{d\phi}+\f{d-1}{2}k^2\f{d\psi(\phi)}{d\phi}-\f{dW(\phi)}{d\phi}\f{U_2}{2U_1}a'^2_t\nn&+&\f{(d-1)B^2}{4h^2}\f{dW(\phi)}{d\phi}\bigg]=0.
\eea
\paragraph{Hawking temperature:} If we construct a black hole solution then the temperature associated to it takes the following form

\be
T_H=\f{1}{4\pi}\left(\f{U'_1\sqrt{U_2}}{\sqrt{U_1}}\right)_{r_h},
\ee
where $r_h$ is the size of the horizon.

The Bekenstein-Hawking entropy density is
\be
s=\f{h^{\f{d-1}{2}}(r_h)}{4G}.
\ee

\subsection{Fluctuation and equations}
Let us fluctuate the background geometry and the matter as
\be
g_{MN}\longrightarrow g^{(0)}_{MN}+H_{MN},\quad F_{MN}\longrightarrow F^{(0)}_{MN}+f_{MN},\quad \chi_i\longrightarrow \chi^{(0)}_i+\delta\chi_i,
\ee 
where the superscript $(0)$ in each field denotes background solution and the other  part corresponds to fluctuation. Under such a fluctuation, the Ricci tensor changes to leading order in fluctuation as 
\be
{\cal R}_{MN}\longrightarrow {\cal R}^{(0)}_{MN}+ R^{(1)}_{MN},
\ee
where 
\be
 R^{(1)}_{MN}=\f{1}{2}\left[ \nabla^{(0)}_K\nabla^{(0)}_N {H^K}_M+\nabla^{(0)}_K\nabla^{(0)}_M {H^K}_N-\nabla^{(0)}_K\nabla^{(0)K} H_{MN}-g^{(0){KL}}\nabla^{(0)}_M\nabla^{(0)}_N H_{KL}\right].
\ee
The covariant derivatives, $\nabla^{(0)}_K$,  are defined with respect to the unperturbed metric $g^{(0)}_{KL}$.

The resulting fluctuating equation of motions associated to geometry and gauge fields are 
\bea
R^{(1)}_{MN}&-&\f{W}{2}\left[ F^{(0)}_{MK} F^{(0)}_{NL} g^{(0)KR}H_{RS}g^{(0)SL}+F^{(0)}_{MK} f_{NL} g^{(0)KL}+F^{(0)}_{NL} f_{MK} g^{(0)KL}\right]-H_{MN}\f{(V+2\Lambda)}{(d-1)}\nn&+&
\f{W}{4(d-1)}H_{MN} F^{(0)KL} F^{(0)}_{KL}+\f{W}{2(d-1)}g^{(0)}_{MN}\left[F^{(0)KL} f_{KL}-{F^{(0)R}}_L F^{(0)SL}H_{RS} \right]\nn&-&
\f{\psi}{2}\p_M \chi^{(0)}_i\p_N\delta\chi_i-\f{\psi}{2}\p_M\delta\chi_i\p_N \chi^{(0)}_i=0,\nn
&&\p_K\Bigg[\sqrt{-g^{(0)}}W\bigg(\left[H^{MK}g^{(0)NL}-H^{ML}g^{(0)NK}\right]F^{(0)}_{MN}+f_{MN}g^{(0)MK}g^{(0)NL}+\nn&&
\f{1}{2}g^{(0)RS}H_{RS}F^{(0)}_{MN}g^{(0)MK}g^{(0)NL}\bigg)\Bigg]=0.
\eea
The superscript indices on metric fluctuations are defined as $H^{MN}\equiv-g^{(0)MK}H_{KL}g^{(0)LN}$.

\paragraph{Traceless fluctuation:} For our purpose, we shall consider only traceless fluctuation to geometry
($g^{(0)RS}H_{RS}=0$), which  can be written    in the  following way
\bea\label{fluctuation_geometry_without_b}
ds^2&=&-U_1(r)dt^2+\f{1}{U_2(r)}dr^2+h(r)\left( dx^2_1+\cdots+dx^2_{d-1}\right)+2G_{t {i}}(t, r)dt dx^i
\nn&+&
2G_{ri}(r)dx^i dr,\quad \chi_i=k~~x^i+\delta\chi_i(r), \nn F&=&a'_t(r)dr\w dt+B\left( dx_1\w dx_2+dx_3\w dx_4+\cdots+dx_{d-2}\w dx_{d-1}\right)\nn&+&\p_ta_{{i}}(t,r)dt\w dx^i+\p_ra_{{i}}(t,r)dr\w dx^i,
\eea
where the fluctuating part of the metric and other fields shall be considered infinitesimally. We shall consider the fluctuation to have the following structure \cite{Banks:2015wha}
\bea\label{flu_metric_gauge_withgout_b}
a_{{i}}(t,r)&=&-E_{{i}} t+a_{{i}}(r)+\xi_{{i}} ~t~ a_t(r),\nn
G_{t{{i}}}(t,r)&=&h(r) h_{{{ti}}}(r)-t~ \xi_{{i}}~  U_1(r),\nn
G_{r{{i}}}&=&h(r) h_{{{i}}}(r),
\eea
where $\xi_i$'s  are related to  thermal gradients along the spatial directions. The advantage of using  such a traceless perturbation is that it is easier to decouple the fluctuating metric components  from the rest.
With such  form of the fluctuation of the geometry and matter fields, the non-zero Ricci tensor components  $R^{(1)}_{t{i}}$ and $R^{(1)}_{r{i}}$ reads as
\bea
R^{(1)}_{t{i}}&=&-\f{1}{2}h U_2 h''_{t{i}}+h'_{t{i}}\left[\f{hU_2 U'_1}{4U_1}-\f{(d+1)}{4}U_2h'-\f{1}{4}hU'_2\right]\nn&-&h_{t{i}}\left(\f{h'U_2 U'_1}{4U_1}+\f{(d-3)}{4h}U_2h'^2+\f{1}{4}h'U'_2+\f{1}{2}U_2 h''\right)\nn&+&t \xi_{{i}}\left[\f{(d-1)}{4h}U_2h'U'_1-\f{U_2U'^2_1}{4U_1}+\f{1}{4}U'_1U'_2+\f{1}{2}U_2U''_1\right],\nn
R^{(1)}_{r{i}}&=&-h_{r{i}}\left[\f{(d-3)U_2 h'^2}{4h}+\f{U_2h'U'_1}{4U_1}+\f{h'U'_2}{4}+\f{U_2h''}{2}\right]+\xi_{{i}}\left[\f{U'_1}{2U_1}-\f{h'}{2h}\right].
\eea

This results in the equation of motion for the $h_{t{1}}$ and $h_{t{2}}$ as

\bea\label{eom_flu_I}
&&-\f{1}{2}h U_2 h''_{t{1}}-h'_{t{1}}\Bigg[\f{B^2 W+2k^2 h \psi+3d U_2 h'^2+2h h'U'_2}{8h'}+\nn&&\f{h^2}{4(d-1) U_1 h'}\Bigg(2U_1(V+2\Lambda)+U_2(Wa'^2_t-U_1\phi'^2)\Bigg)\Bigg]+h_{t{1}}\Bigg[ \f{B^2 W}{2h}+\f{k^2\psi}{2}\Bigg]\nn&+&\f{BW}{2h}(E_{{2}}-\xi_{{2}} a_t)-\f{1}{2}BU_2 W h_{r{{2}}}a'_t-\f{1}{2}U_2 Wa'_t a'_{{1}}=0,\nn
&&-\f{1}{2}h U_2 h''_{t{2}}-h'_{t{2}}\Bigg[\f{B^2 W+2k^2 h \psi+3d U_2 h'^2+2h h'U'_2}{8h'}+\nn&&\f{h^2}{4(d-1) U_1 h'}\Bigg(2U_1(V+2\Lambda)+U_2(Wa'^2_t-U_1\phi'^2)\Bigg)\Bigg]+h_{t{2}}\Bigg[ \f{B^2 W}{2h}+\f{k^2\psi}{2}\Bigg]\nn&-&\f{BW}{2h}(E_{{1}}-\xi_{{1}} a_t)+\f{1}{2}BU_2 W h_{r{{1}}}a'_t-\f{1}{2}U_2 Wa'_t a'_{{2}}=0,
\eea

where we have used eq(\ref{eom}). The equation of motion associated to the other components of the metric fluctuation  simply follows by looking at eq(\ref{eom_flu_I}). As an example, the equation of motion for $h_{t{3}}$ follows from the first equation of eq(\ref{eom_flu_I}) with the following substitution: 
\be\label{pres_1}
h_{t{1}}\ra h_{t{3}},\quad (E_{{2}},\xi_{{2}})\ra (E_{{4}},\xi_{{4}}), \quad h_{r{{2}}}\ra h_{r{{4}}}.
\ee

Similarly, the equation of motion for $h_{t{4}}$ follows from the second equation of eq(\ref{eom_flu}) with the following substitution: 
\be\label{pres_2}
h_{t{2}}\ra h_{t{4}},\quad (E_{{1}},\xi_{{1}})\ra (E_{{3}},\xi_{{3}}), \quad h_{r{{1}}}\ra h_{r{{3}}}.
\ee
Most importantly, $h_{t{1}},~h_{t{2}},~h_{r{1}}$ and $h_{r{2}}~$ decouples from the rest of the metric fluctuations.

The equation of motion of other components of the fluctuating metric $h_{t{i}}$ follows, similarly. The equation of motion of the fluctuating gauge field reads as
\bea
\p_r\left[ \sqrt{U_1U_2} W h^{\f{d-3}{2}}\left( a'_i\delta^{in}-Bh_{ri}\delta^{ij}\delta^{mn}\epsilon_{jm}+\f{U_2}{U_1}a'_t\delta^{in}h_{ti}\right)\right]+\sqrt{\f{U_1}{U_2}} W h^{\f{d-5}{2}}B\delta^{ij}\xi_i\delta^{mn}\epsilon_{jm}=0.
\eea
\subsection{Currents}

\paragraph{Electric currents:}The radially conserved electric currents for the Einstein-Maxwell-dilaton-axion system are

\bea\label{conserved_electric_current}
J^{1}(r)&=&-\f{\sqrt{U_1U_2}}{(16\pi G) } W h^{\f{d-3}{2}}\left[ a'_{{1}}+B  h_{r{{2}}}+\f{h}{U_1}a'_t h_{t{1}}\right]-\f{\xi_{{2}}}{(16\pi G) } M_J(r),\quad \nn
J^{2}(r)&=&-\f{\sqrt{U_1U_2}}{(16\pi G) } W h^{\f{d-3}{2}}\left[ a'_{{2}}-B  h_{r{{1}}}+\f{h}{U_1}a'_t h_{t{2}}\right]+\f{\xi_{1}}{(16\pi G)} M_J(r),\nn
M_J(r)&=&-B \int_{r_h}^{r} \sqrt{\f{U_1}{U_2}}Wh^{\f{d-5}{2}},
%\quad M^2(r)=-B \xi_{{1}}\int_{r_h}^{r} \sqrt{\f{U_1}{U_2}}Wh^{\f{d-5}{2}},
\eea

The other components of the radially conserved currents can be written by following the  prescription as written down in eq(\ref{pres_1}) and eq(\ref{pres_2}).

\paragraph{Heat currents:} Let us consider  quantities, ${\cal Q}^1(r)$ and ${\cal Q}^2(r)$, which has the structure as follows
\bea
{\cal Q}^1(r)&=&\f{U^{\f{3}{2}}_1\sqrt{U_2}}{(16\pi G) }  h^{\f{d-3}{2}}\p_r\left( \f{h h_{t1}}{U_1} \right)-a_t(r) J^1(r)\nn
{\cal Q}^2(r)&=&\f{U^{\f{3}{2}}_1\sqrt{U_2}}{(16\pi G) }  h^{\f{d-3}{2}}\p_r\left( \f{h h_{t2}}{U_1} \right)-a_t(r) J^2(r),
\eea

It follows that the radial gradient of ${\cal Q}^1(r)$ and ${\cal Q}^2(r)$ can be calculated  using the fluctuating equation of motion for $h_{t1}(r)$ and $h_{t2}(r)$ as written down in eq(\ref{eom_flu_I}) as well as the equation of motion of $U_1(r),~U_2(r)$ and $h(r)$ as written in  eq(\ref{eom}), which results in 
\bea
\p_r {\cal Q}^1&=&\f{BW}{16\pi G}\sqrt{\f{U_1}{U_2}}h^{\f{d-5}{2}}(E_{{2}}-\xi_{{2}} A_t)+\f{\xi_2}{16\pi G}M_J(r)a'_t(r)\equiv {\cal M}^1(r),\nn
\p_r {\cal Q}^2&=&-\f{BW}{16\pi G}\sqrt{\f{U_1}{U_2}}h^{\f{d-5}{2}}(E_{{1}}-\xi_{{1}} A_t)-\f{\xi_1}{16\pi G}M_J(r)a'_t(r)\equiv {\cal M}^2(r).
\eea

The radially conserved heat currents can be constructed as follows
\be
Q^1(r)={\cal Q}^1(r)-\int^r_{r_h} dx {\cal M}^1(x),\quad Q^2(r)={\cal Q}^2(r)-\int^r_{r_h} dx {\cal M}^2(x).
\ee

One can similarly construct the other  radially conserved heat currents along the other spatial directions.

\subsubsection{Currents at the horizon}

In order to calculate the currents at the horizon, we need to find out the behavior of fields at the horizon. Essentially, we want to put the in-falling boundary conditions at  the horizon and  are as follows: 

\bea
a_1(r)&=&-\f{E_1}{U_0}~Log(r-r_h)+{\cal O}(r-r_h),\quad a_2(r)=-\f{E_2}{U_0}~Log(r-r_h)+{\cal O}(r-r_h),\nn
h_{t1}(r)&=& Uh_{r1}(r_h)-\xi_1 \left(\f{U_1(r)}{h(r)~ U_0}\right)~Log(r-r_h)+{\cal O}(r-r_h), \nn 
h_{t2}(r)&=&Uh_{r2}(r_h)-\xi_2\left(\f{U_1(r)}{h(r)~ U_0}\right)~Log(r-r_h)+{\cal O}(r-r_h),
\eea
where $U(r)\equiv \sqrt{U_1(r)U_2(r)}=U_0 (r-r_h)+\cdots$. Note, $U_0$ is independent of $r$ and respects the relation  $U_0=\sqrt{U^{(0)}_1U^{(0)}_2}$ as  we demand that the function $U_1(r)$ and $U_2(r)$ at the horizon  has the form as follows:
\be\label{u1_u2_horizon}
U_1(r)=U^{(0)}_1(r-r_h)+{\cal O}(r-r_h)^2,\quad  U_2(r)=U^{(0)}_2(r-r_h)+{\cal O}(r-r_h)^2.
\ee
This allows us to write he temperature as
\be
T_H=\f{\sqrt{U^{(0}_1U^{(0)}_2}}{4\pi}\equiv \f{U_0}{4\pi},
\ee

\paragraph{Electric currents at the horizon:} The currents at the horizon with the help of the in-falling boundary condition  reads as 

\bea
(16\pi G) J^{1}(r_h)&=&\left[W h^{\f{d-3}{2}} \bigg( E_1-B  h_{t{{2}}}-\f{\rho }{Wh^{\f{d-3}{2}}} h_{t{1}}\bigg)\right]_{r_h}\nn
(16\pi G) J^{2}(r_h)&=&\left[W h^{\f{d-3}{2}} \bigg( E_{{2}}+B  h_{t{{1}}}-\f{\rho }{Wh^{\f{d-3}{2}}} h_{t{2}}\bigg)\right]_{r_h},\\
(16\pi G)~        Q^1(r_h)&=&-U_0h^{\f{d-1}{2}}(r_h)h_{t1}(r_h), \quad 
(16\pi G)~ Q^2(r_h)=-U_0h^{\f{d-1}{2}}(r_h)h_{t2}(r_h).
\eea

In order to calculate the currents at the horizon in terms of the electric fields and the thermal gradients, we need to know the behavior of $h_{t1}$ at the horizon. It can be calculated with the help of  eq(\ref{constraint}) and the first equation of eq(\ref{eom_flu_I}). This resulted in the following fluctuating equation

\bea\label{eom_flu}
&&-\f{1}{2}h U_2 h''_{t{1}}-h'_{t{1}}\Bigg[\f{(d+1) U_1U_2h'-hU_2U'_1+hU_1U'_2}{4U_1}\Bigg]+h_{t{1}}\Bigg[ \f{B^2 W}{2h}+\f{k^2\psi}{2}\Bigg]\nn&+
&\f{BW}{2h}(E_{{2}}-\xi_{{2}} a_t)-\f{1}{2}BU_2 W h_{r{{2}}}a'_t-\f{1}{2}U_2 Wa'_t a'_{{1}}=0
\eea

Upon inspection, the  behavior of this differential equation with the help of the in-falling boundary condition at the horizon gives
\be\label{coupled_ht1_ht2_I}
\Bigg[\f{\xi_1}{2} U_0+h_{t{1}}\left( \f{B^2 W}{2h}+\f{k^2\psi}{2}\right)+\f{BW}{2h}E_{{2}}-\f{B \rho}{2h^{\f{d-1}{2}}} h_{t2}
+\f{E_1 \rho}{2h^{\f{d-1}{2}}}\Bigg]_{r_h} =0
\ee

Similarly, the differential equation obeyed by $h_{t2}(r)$ is 

\bea
&&-\f{1}{2}h U_2 h''_{t{2}}-h'_{t{2}}\Bigg[\f{(d+1)U_1 U_2 h'+h(U_1 U'_2-U_2U'_1)}{4U_1}\Bigg] +h_{t{2}}\Bigg[ \f{B^2 W}{2h}+\f{k^2\psi}{2}\Bigg]\nn&-&\f{BW}{2h}(E_{{1}}-\xi_{{1}} a_t)+\f{1}{2}BU_2 W h_{r{{1}}}a'_t-\f{1}{2}U_2 Wa'_t a'_{{2}}=0,
\eea

 The   behavior of this differential equation with the help of the in-falling boundary condition at the horizon gives
 
 \be\label{coupled_ht1_ht2_II}
 \Bigg[\f{\xi_2}{2} U_0+h_{t{2}}\left( \f{B^2 W}{2h}+\f{k^2\psi}{2}\right)-\f{BW}{2h}E_{{1}}+\f{B \rho}{2h^{\f{d-1}{2}}} h_{t1}
 +\f{E_2 \rho}{2h^{\f{d-1}{2}}}\Bigg]_{r_h} =0
 \ee

We can solve for the equations eq(\ref{coupled_ht1_ht2_I}) and eq(\ref{coupled_ht1_ht2_II}) and the behavior of $h_{t1}(r_h)$ and $h_{t1}(r_h)$ at the horizon reads as
\bea
h_{t1}(r_h)&=&-\Bigg[\f{  1}{ B^4h^d W^2+h^{ d+2} k^4 \psi^2 + 
	B^2 h^3 ( \rho^2+ 2 h^{d-2} k^2 W \psi)}\times\nn&&[E_1 h^{( d+5)/2} k^2 \rho \psi+ B E_2h^3 ( \rho^2 +B^2  h^{d-3} W^2+ h^{d-2} k^2 W \psi)+\nn&&\xi_1U_0h^{d+1}(B^2W+hk^2\psi)+\xi_2BU_0\rho h^{( d+5)/2}]\Bigg]_{r_h},\nn
h_{t2}(r_h)&=&\Bigg[\f{  1}{ B^4h^d W^2+h^{ d+2} k^4 \psi^2 + 
	B^2 h^3 ( \rho^2+ 2 h^{d-2} k^2 W \psi)}\times\nn&&[B E_1h^3 ( \rho^2 +B^2  h^{d-3} W^2+ h^{d-2} k^2 W \psi)+E_2 h^{( d+5)/2} k^2 \rho \psi +\nn&&\xi_1BU_0\rho h^{( d+5)/2} +\xi_2U_0h^{d+1}(B^2W+hk^2\psi)]\Bigg]_{r_h}.
\eea

\section{Transport quantities}
The electric currents and the heat currents at the horizon can be written in terms of the electrical conductivity, thermo-electric and thermal conductivity as follows

\bea\label{electrical_heat_current}
J^1(r_h)&=&\sigma_{11}(r_h)E_1+\sigma_{12}(r_h)E_2+T_H\alpha_{11}(r_h)\xi_1+T_H\alpha_{12}(r_h)\xi_2,\nn
J^2(r_h)&=&\sigma_{21}(r_h)E_1+\sigma_{22}(r_h)E_2+T_H\alpha_{21}(r_h)\xi_1+T_H\alpha_{22}(r_h)\xi_2,\nn
 Q^1(r_h)&=&T_H{\overline\alpha}_{11}(r_h)E_1+T_H{\overline\alpha}_{12}(r_h)E_2+T_H{\overline\kappa}_{11}(r_h)\xi_1+T_H{\overline\kappa}_{12}(r_h)\xi_2,\nn
 Q^2(r_h)&=&T_H{\overline\alpha}_{21}(r_h)E_1+T_H{\overline\alpha}_{22}(r_h)E_2+T_H{\overline\kappa}_{21}(r_h)\xi_1+T_H{\overline\kappa}_{22}(r_h)\xi_2.
\eea

Similar expressions exists for other currents. The transport coefficients takes the following form
\bea\label{transport_b}
\sigma_{11}(r_h)&=&\sigma_{22}(r_h)=\f{1}{16\pi G}\left[\psi k^2h^{\f{d-3}{2}}\f{(\rho^2 h^{2-d}+W\psi k^2+W^2B^2h^{-1})}{\psi^2 k^4+B^2h^{-1}(\rho^2 h^{2-d}+2W\psi k^2+W^2B^2h^{-1})}\right]_{r_h}\nn
\sigma_{12}(r_h)&=&-\sigma_{21}(r_h)=\f{1}{16\pi G}\left[\rho Bh^{-1}\f{(\rho^2 h^{2-d}+2W\psi k^2+W^2B^2h^{-1})}{\psi^2 k^4+B^2h^{-1}(\rho^2 h^{2-d}+2W\psi k^2+W^2B^2h^{-1})}\right]_{r_h}\nn
\alpha_{11}(r_h)&=&\alpha_{22}(r_h)={\overline\alpha}_{11}(r_h)={\overline\alpha}_{22}(r_h)\nn&=& \f{1}{16\pi G}\left[\f{4\pi \rho\psi k^2}{\psi^2 k^4+B^2h^{-1}(\rho^2 h^{2-d}+2W\psi k^2+W^2B^2h^{-1})}\right]_{r_h}\nn
\alpha_{12}(r_h)&=&-\alpha_{21}(r_h)={\overline\alpha}_{12}(r_h)=-{\overline\alpha}_{21}(r_h)\nn&=& \f{1}{16\pi G}\left[4\pi B h^{\f{d-3}{2}}\f{(\rho^2 h^{2-d}+W\psi k^2+W^2B^2h^{-1})}{\psi^2 k^4+B^2h^{-1}(\rho^2 h^{2-d}+2W\psi k^2+W^2B^2h^{-1})}\right]_{r_h}\nn
{\overline\kappa}_{11}(r_h)&=&{\overline\kappa}_{22}(r_h)=\f{1}{16\pi G}\left[16\pi^2 T_H h^{\f{d-1}{2}}\f{(\psi k^2+WB^2h^{-1})}{\psi^2 k^4+B^2h^{-1}(\rho^2 h^{2-d}+2W\psi k^2+W^2B^2h^{-1})}\right]_{r_h}\nn
{\overline\kappa}_{12}(r_h)&=&-{\overline\kappa}_{21}(r_h)=\f{1}{16\pi G} \left[\f{16\pi^2 T_H\rho B}{\psi^2 k^4+B^2h^{-1}(\rho^2 h^{2-d}+2W\psi k^2+W^2B^2h^{-1})}\right]_{r_h}\nn
\eea 

The thermal conductivity is defined as the response of the heat current  to the  thermal gradient for zero electric current,  $\kappa_{ij}={\overline\kappa}_{ij}-T_H(\alpha\sigma^{-1}\alpha)_{ij}$

\bea\label{transport_kappa}
\kappa_{11}(r_h)&=&\kappa_{22}(r_h)=\Bigg[{\overline\kappa}_{11}-T_H\f{(\alpha^2_{11}-\alpha^2_{12})\sigma_{11}+2\alpha_{11}\alpha_{12}\sigma_{12}}{\sigma^2_{11}+\sigma^2_{12}}\Biggr]_{r_h}\nn&=&\f{1}{16\pi G}\left[16\pi^2 T_H W h^{\f{d-1}{2}}\f{(\rho^2 h^{2-d}+W\psi k^2)}{B^2 W^2\rho^2 h^{1-d} +(\rho^2 h^{2-d}+W\psi k^2)^2}\right]_{r_h}\nn
\kappa_{12}(r_h)&=&-\kappa_{21}(r_h)=\Biggl[{\overline\kappa}_{12}+T_H\f{(\alpha^2_{11}-\alpha^2_{12})\sigma_{12}-2\alpha_{11}\alpha_{12}\sigma_{11}}{\sigma^2_{11}+\sigma^2_{12}}\Biggr]_{r_h}\nn&=&-\f{1}{16\pi G}\left[  \f{16\pi^2 B T_HW^2\rho}{B^2 W^2\rho^2 h^{1-d} +(\rho^2 h^{2-d}+W\psi k^2)^2}\right]_{r_h}.
\eea
Interestingly, the Nerst coefficient reads as
\be
\nu\equiv \f{\alpha_{12}(r_h)}{B\sigma_{11}(r_h)}=\f{4\pi}{\psi k^2}.
\ee

The Seebeck coefficient and the Nernst response is defined as ratio of the electric field to the thermal gradient in the absence of electrical current, $\vartheta=-\sigma^{-1}\cdot \alpha.$ The  Seebeck coefficient $\vartheta_{11}$ and the Nernst response $\vartheta_{12}$ reads as
\bea\label{nernst_coefficient}
\vartheta_{11}(r_h)&=&-\f{\sigma_{11}(r_h)\alpha_{11}(r_h)+\sigma_{12}(r_h)\alpha_{12}(r_h)}{\sigma^2_{11}(r_h)+\sigma^2_{12}(r_h)}=-4\pi\rho\left[\f{h^{\f{3-d}{2}}(\rho^2 h^{2-d}+W\psi k^2+W^2B^2h^{-1})}{B^2 W^2\rho^2 h^{1-d} +(\rho^2 h^{2-d}+W\psi k^2)^2}\right]_{r_h},\nn
\vartheta_{12}(r_h)&=&-\f{\sigma_{11}(r_h)\alpha_{12}(r_h)-\sigma_{12}(r_h)\alpha_{11}(r_h)}{\sigma^2_{11}(r_h)+\sigma^2_{12}(r_h)}=-4\pi B\left[\f{W^2\psi k^2}{B^2 W^2\rho^2 h^{1-d} +(\rho^2 h^{2-d}+W\psi k^2)^2}\right]_{r_h}
\eea
Let us also mention the resistivity matrix
\bea\label{resistivity}
\rho_{11}(r_h)&\equiv&\f{\sigma_{11}(r_h)}{\sigma^2_{11}(r_h)+\sigma^2_{12}(r_h)}=(16\pi G)\left[\f{h^{\f{3-d}{2}}\psi k^2(\rho^2 h^{2-d}+W\psi k^2+W^2B^2h^{-1})}{B^2 W^2\rho^2 h^{1-d} +(\rho^2 h^{2-d}+W\psi k^2)^2}\right]_{r_h}\nn&=&-(16\pi G)\left[\f{\psi k^2}{4\pi\rho}\vartheta_{11}(r)\right]_{r_h},\nn
\rho_{12}(r_h)&\equiv&-\f{\sigma_{12}(r_h)}{\sigma^2_{11}(r_h)+\sigma^2_{12}(r_h)}=-(16\pi G)\rho B\left[\f{h^{2-d}(\rho^2 h^{2-d}+W\psi k^2+W^2B^2h^{-1})}{B^2 W^2\rho^2 h^{1-d} +(\rho^2 h^{2-d}+W\psi k^2)^2}\right]_{r_h}\nn&=&(16\pi G)\left[\f{Bh^{\f{1-d}{2}}}{4\pi}\vartheta_{11}(r)\right]_{r_h}.\nn
\eea
It simply follows that 
\be\label{thetaoverrestivity}
\f{\vartheta_{11}(r_h)}{\rho_{11}(r_h)}=-\f{1}{(16\pi G)}\left(\f{4\pi\rho}{\psi(\phi(r_h)) k^2}\right),\quad \f{\vartheta_{12}(r_h)}{\kappa_{12}(r_h)}=\f{16\pi G}{T_H}\left(\f{\psi(\phi(r_h)) k^2}{4\pi\rho}\right).
\ee
\section{A  transformation}
Let us consider a transformation, inspired from  \cite{Hartnoll:2007ih} for $d=3$, in  which the charge density and the magnetic field gets interchanged. For generic $d$, the transformation reads as 
\bea\label{duality}
\rho\ra B~~  W(\phi(r_h))  ~~ h^{\f{d-3}{2}}(r_h),\quad B \ra \f{\rho}{W(\phi(r_h))~~  h^{\f{d-3}{2}}(r_h)}.
\eea
It is easy to see that under such a transformation the quantity $\rho B$ as well as $B^2 W^2 h^{-1} +\rho^2 h^{2-d}$ evaluated at the horizon remains invariant whereas $\rho/B\ra (B/\rho) W^2 h^{d-3}$. This transformation is interpreted in \cite{Hartnoll:2007ih} as particle vortex duality.

Under such a  transformation, eq(\ref{duality}), the transport coefficients transforms as
\bea
&&\sigma_{11}(r_h)\leftrightarrow \f{\rho_{11}(r_h)}{(16 \pi G)^2} ~~(W(\phi(r_h)) )^2 ~~ h^{d-3}(r_h), \quad \sigma_{12}(r_h)\leftrightarrow -\f{\rho_{12}(r_h)}{(16 \pi G)^2} ~~(W(\phi(r_h)) )^2 ~~ h^{d-3}(r_h),\nn
&&\alpha_{11}(r_h)\leftrightarrow -\f{\vartheta_{12}(r_h)}{(16 \pi G)} ~~W(\phi(r_h))  ~~ h^{\f{d-3}{2}}(r_h), \quad \alpha_{12}(r_h)\leftrightarrow -\f{\vartheta_{11}(r_h)}{(16 \pi G)} ~~W(\phi(r_h))  ~~ h^{\f{d-3}{2}}(r_h),\nn
&&{\overline\kappa}_{11}(r_h)\leftrightarrow \kappa_{11}(r_h), \quad {\overline\kappa}_{12}(r_h)\leftrightarrow -\kappa_{12}(r_h)
\eea

If we impose of  either the condition $\sigma_{11}(r_h)=\f{\rho_{11}(r_h)}{(16 \pi G)^2}(W(\phi(r_h)) )^2 ~ h^{d-3}(r_h)$ or $\sigma_{12}(r_h)= -\f{\rho_{12}(r_h)}{(16 \pi G)^2} ~(W(\phi(r_h)) )^2 ~ h^{d-3}(r_h)$  then it provides an interesting relation among the  electrical conductivities
\be\label{circle}
\sigma^2_{11}(r_h)+\sigma^2_{12}(r_h)=(W(\phi(r_h)) )^2 ~ h^{d-3}(r_h)
\ee

It is an equation of a circle and 
this relation holds for non-zero values of electrical conductivities. More importantly, it relates the sum of squares of  electrical conductivities  with the geometrical quantities. If we demand that the longitudinal electrical conductivity as well as the transverse electrical conductivity remain positive, which happens for positive values of $\psi$,  then eq(\ref{circle}) will be an equation for a semi-circle.

The meaning of the condition $\sigma_{11}(r_h)=\f{\rho_{11}(r_h)}{(16 \pi G)^2}(W(\phi(r_h)) )^2 ~ h^{d-3}(r_h)$ or \\ $\sigma_{12}(r_h) = -\f{\rho_{12}(r_h)}{(16 \pi G)^2} ~(W(\phi(r_h)) )^2 ~ h^{d-3}(r_h)$ essentially relates the charge density, magnetic field and the geometric quantities as
 $\rho=B~~  W(\phi(r_h))  ~~ h^{\f{d-3}{2}}(r_h) $. 

\paragraph{The special point:} At  $\rho=B~~  W(\phi(r_h))  ~~ h^{\f{d-3}{2}}(r_h) $, we can show that as long as $W(\phi(r_h))$ and $\psi(\phi(r_h))$ are positive the following relation holds
\bea
&&\sigma^2_{11}(r_h)<  \sigma^2_{12}(r_h)\quad {\rm for}\quad h(r_h)\psi(\phi(r_h)) k^2 < \sqrt{2} W(\phi(r_h)) B^2\nn
&&\alpha^2_{11}(r_h) < \alpha^2_{12}(r_h)\quad {\rm generically}\nn
&&\kappa^2_{11}(r_h) > \kappa^2_{12}(r_h)\quad {\rm generically}.
\eea 
We can also show 
\be
\f{\alpha^2_{11}(r_h)-\alpha^2_{12}(r_h)}{\alpha^2_{11}(r_h)\alpha_{12}(r_h)}=-(16\pi G)\left(2\f{\sigma_{12}(r_h)}{\sigma_{11}(r_h)}\right),\quad \kappa_{12}(r_h)=-T_H\f{\alpha_{11}(r_h)\alpha_{12}(r_h)}{\sigma_{11}(r_h)}.
\ee

\subsection{Universal relations} After closer inspection of the transport coefficients as written in eq(\ref{transport_b}) reveals an interesting universal relation among the different components of the transport quantities.  
It follows that the ratio of specific transport coefficients are
\be\label{universal_ratio}
\f{\alpha_{12}(r_h)}{\sigma_{11}(r_h)}=\left(\f{4\pi B}{\psi k^2}\right)_{r_h},\quad T_H\f{\alpha_{11}(r_h)}{\overline\kappa_{12}(r_h)}=\left(\f{\psi k^2}{4\pi B}\right)_{r_h}
\ee

Upon combining these relations, we get

\be\label{universal_relation}
   T_H\f{\alpha_{11}(r_h)}{\overline\kappa_{12}(r_h)}\f{\alpha_{12}(r_h)}{\sigma_{11}(r_h)}=1.
\ee
It is universal in the sense that it does not depend on the number of spacetime dimensions ($d\geq 3$). However, for these relations to work we need to have  non-zero magnetic field. 

The transport coefficients for the Einstein-DBI-dilaton-axion system for $d=3$ has been calculated in \cite{Pal:2019bfw} and \cite{Pal:2020gsq}.  
The relation among the transport coefficients, i.e.,  eq(\ref{universal_relation}) is respected  for the DBI case too. It essentially suggests us to make a conjecture that the relation eq(\ref{universal_relation}) does not depend on the nature of the matter field. However, this claim need to be checked by looking at other systems.

The universal relation as written down in eq(\ref{universal_relation}) can be re-written as 
\bea\label{int_relation}
  &&T_H\left[ \sigma_{11}(r_h)\sigma_{12}(r_h)(\alpha^2_{11}(r_h)-\alpha^2_{12}(r_h))-\alpha_{11}(r_h)\alpha_{12}(r_h)(\sigma^2_{11}(r_h)-\sigma^2_{12}(r_h))\right]\nn&=&\sigma_{11}(r_h)\kappa_{12}(r_h)(\sigma^2_{11}(r_h)+\sigma^2_{12}(r_h)).
\eea

It essentially says that the transport coefficient, $\kappa_{12}(r_h)$, is not an independent quantity rather is completely determined by the electrical and thermo-electric coefficients.  Moreover, the  relation eq(\ref{universal_relation}) or  eq(\ref{int_relation}) does not depend on explicitly on either the nature of the geometry or the form of couplings, even though each transport coefficients does.

The quantity $\f{\alpha_{11}(r_h)}{\overline\kappa_{12}(r_h)}\f{\alpha_{12}(r_h)}{\sigma_{11}(r_h)}$ under the transformation eq(\ref{duality}) transforms as
\be
\f{\alpha_{11}(r_h)}{\overline\kappa_{12}(r_h)}\f{\alpha_{12}(r_h)}{\sigma_{11}(r_h)}\ra -\f{\vartheta_{12}(r_h)}{\kappa_{12}(r_h)}\f{\vartheta_{11}(r_h)}{\rho_{11}(r_h)}=\f{1}{T_H},
\ee
where in the last equality, we have used the value of transport coefficients evaluated in eq(\ref{transport_kappa}), eq(\ref{nernst_coefficient}) and eq(\ref{resistivity}). Hence, it follows that the relation eq(\ref{universal_relation}) predicts another relation under eq(\ref{duality})
\be
-T_H\f{\vartheta_{12}(r_h)}{\kappa_{12}(r_h)}\f{\vartheta_{11}(r_h)}{\rho_{11}(r_h)}=1
\ee
This is in agreement with the second equation of eq(\ref{thetaoverrestivity}).
There also exists  interesting relations that follows  eq(\ref{thetaoverrestivity}) and  eq(\ref{universal_ratio}) and  reads as
\bea
&&\f{\sigma_{11}(r_h)}{\alpha_{12}(r_h)}\f{\vartheta_{11}(r_h)}{\rho_{11}(r_h)}=-\f{1}{16\pi G}\left(\f{\rho}{B}\right)=T_H\f{\vartheta_{11}(r_h)}{\rho_{11}(r_h)}\f{\alpha_{11}(r_h)}{\overline\kappa_{12}(r_h)},\nn
&&T_H\f{\alpha_{12}(r_h)}{\sigma_{11}(r_h)}\f{\vartheta_{12}(r_h)}{\kappa_{12}(r_h)}=16\pi G\left(\f{B}{\rho}\right)=\f{\overline\kappa_{12}(r_h)}{\alpha_{11}(r_h)}\f{\vartheta_{12}(r_h)}{\kappa_{12}(r_h)}.
\eea

The electric charge is defined as 
\be
Q=\int d^{d-1}x J^0, \quad{\rm where}\quad  J^0=-\f{W(\phi)}{16\pi G}\sqrt{-g}F^{r0}=\f{\rho}{16\pi G},
\ee
where we have used  the solution to gauge potential. Denoting $\int d^{d-1}x\equiv V_{d-1} $, gives $Q=\f{V_{d-1}\rho}{16\pi G}$. This results in 

\bea
&&\f{\sigma_{11}(r_h)}{\alpha_{12}(r_h)}\f{\vartheta_{11}(r_h)}{\rho_{11}(r_h)}=-\f{1}{V_{d-1}}\left(\f{Q}{B}\right)=T_H\f{\vartheta_{11}(r_h)}{\rho_{11}(r_h)}\f{\alpha_{11}(r_h)}{\overline\kappa_{12}(r_h)},\nn
&&T_H\f{\alpha_{12}(r_h)}{\sigma_{11}(r_h)}\f{\vartheta_{12}(r_h)}{\kappa_{12}(r_h)}=V_{d-1}\left(\f{B}{Q}\right)=\f{\overline\kappa_{12}(r_h)}{\alpha_{11}(r_h)}\f{\vartheta_{12}(r_h)}{\kappa_{12}(r_h)}.
\eea

\section{Special case: At UV}

\paragraph{An exact solution: } The differential equation obeyed by the metric components eq(\ref{eom}) and the dilaton equation eq(\ref{scalar_eom}) can be  solved and the solution reads as 
\bea\label{ads_spacetime}
ds^2_{d+1}&=&\f{r^2}{L^2}\left[-f(r)dt^2+
dx^2_1+\cdots+dx^2_{d-1}\right]+\f{L^2}{r^2f(r)}dr^2,\quad \phi(r)={\rm constant},\nn 
W&=&w_0,\quad \psi=\psi_0,\quad V=0,\quad A'_t(r)=\f{L^{d-1}\rho}{w_0 r^{d-1}},\nn f(r)&=&-\f{2\Lambda L^2}{d(d-1)}-\f{B^2L^6w_0}{4(d-4)r^4}-\f{k^2L^4\psi_0}{2(d-2)r^2}+\f{c_1}{r^d}+\f{\rho^2 L^{2d}}{2w_0(d-1)(d-2)r^{2(d-1)}},
\eea
where the constant, $c_1$, is related to the mass of the black hole. Upon setting the cosmological constant as $2\Lambda=-\f{d(d-1)}{L^2}$, the temperature of the black hole is related to the size of the horizon as
\be\label{temp_b_charge_density_dissipation}
T_H=\f{dr_h}{4\pi L^2}\left[1-\f{B^2 L^6w_0}{4dr^4_h}-\f{k^2L^4\psi_0}{2dr^2_h}-\f{
	\rho^2L^{2d}}{2w_0d(d-1)r^{2(d-1)}_h} \right].
\ee 
\paragraph{Thermodynamic Quantities:} Various thermodynamic quantities like entropy (S), charge (Q), energy (E), free energy (F), magnetization (m) and magnetic susceptibility ($\chi_B$) are as follows:
\bea\label{td_uv}
S&=&\f{V_{d-1}}{4G}\f{r^{d-1}_h}{L^{d-1}},\quad Q=\f{V_{d-1}}{16\pi G}\rho,\nn
E&=&\f{V_{d-1}}{16\pi G}\left[\f{(d-1)r^d_h}{L^{d+1}} -\f{(d-1)B^2w_0}{4(d-4)}\f{r^{d-4}_h}{L^{d-5}}-\f{(d-1)k^2\psi_0}{2(d-2)}\f{r^{d-2}_h}{L^{d-3}}+\f{\rho^2}{2(d-2)w_0}\f{L^{d-1}}{r^{d-2}_h}\right],\nn
F&=&-\f{V_{d-1}}{16\pi G}\left[\f{r^d_h}{L^{d+1}} +\f{3B^2w_0}{4(d-4)}\f{r^{d-4}_h}{L^{d-5}}+\f{k^2\psi_0}{2(d-2)}\f{r^{d-2}_h}{L^{d-3}}+\f{\rho^2}{2(d-1)(d-2)w_0}\f{L^{d-1}}{r^{d-2}_h}\right],\nn
m&=&-\left(\f{\p F}{\p B}\right)_{\beta} =\f{(d-1)V_{d-1}}{32\pi G(d-4)} B w_0\f{r^{d-4}_h}{L^{d-5}}\quad 
\chi_B(B=0)= \f{(d-1)V_{d-1}}{32\pi G(d-4)} w_0
\left(\f{r^{d-4}_h}{L^{d-5}}\right)_{B=0}
\eea

The energy is calculated by assuming that it obeys the first law of thermodynamics. The thermodynamic quantities  are calculated in the Appendix A by using counter term method for $d=3,~d=5$ and $d=7$. The result matches with that given in eq(\ref{td_uv}).

\paragraph{Transport coefficients with vanishing Magnetic field: }

Without the magnetic fields, the longitudinal transport coefficients takes the following form with the choice $16\pi G=1$

\bea\label{transport_zero_B}
\sigma_{11}(r_h)&=&\sigma_{22}(r_h)=
h^{\f{d-3}{2}}(r_h)\left[W+\f{\rho^2 h^{2-d}}{\psi k^2}\right]_{r_h},\nn
\alpha_{11}(r_h)&=&\alpha_{22}(r_h)=\left[\f{4\pi \rho}{\psi k^2}\right]_{r_h},\nn
\kappa_{11}(r_h)&=&\kappa_{22}(r_h)=16\pi^2 T_H\left[\f{  h^{\f{d-1}{2}}}{\psi k^2(1+\f{\rho^2 h^{2-d}}{W\psi k^2})}\right]_{r_h},\nn
\vartheta_{11}(r_h)&=&-\left[\f{4\pi\rho}{W\psi k^2}\left(\f{h^{\f{3-d}{2}}}{1+\f{\rho^2 h^{2-d}}{W\psi k^2}}\right)\right]_{r_h}.
\eea 

It follows that in the absence of magnetic field, the thermo-electric conductivity is a constant at UV. Moreover, the electrical conductivity for an asymptotically AdS spacetime is computed in \cite{Andrade:2013gsa} without the scalar field $\phi$. Let us compare our results with theirs by setting  $h(r)=r^2/L^2$. Without the scalar field $\phi$, the gauge field can be integrated to give, $A'_t=\f{\rho}{W h^{\f{d-1}{2}}}$.  In the present case, the chemical potential, $\mu$, is related to the charge density as 
\be\label{charge_density_chemical_potential}
\rho L^{d-1}=w_0\mu(d-2)r^{d-2}_h.
\ee 
It means for a fixed charge density, the chemical potential changes with the horizon size in a power law type.
This essentially gives $\rho^2 (h(r_h))^{2-d}L^2=(d-2)^2\mu^2w^2_0$. Hence, the longitudinal electrical conductivity becomes
\be\label{andrade_cond}
\sigma_L\equiv \sigma_{11}(r_h)=\sigma_{22}(r_h)=w_0\left(\f{r_h}{L}\right)^{d-3}\left[1+\f{(d-2)^2w_0\mu^2}{\psi_0k^2L^2}\right]
\ee

This precisely matches with the results reported in  \cite{Andrade:2013gsa} for $w_0=1,~~\psi_0=1$. It also follows that for $d=3$ the longitudinal electrical conductivity is completely determined by the   chemical potential and the momentum dissipation. Recall, from the solution of axion, $\chi$, the quantity $k$ can be interpreted as the source term. Hence, the  longitudinal electrical conductivity for $d=3$ is fully determined by the source term of the gauge potential and the axion. The thermo-electric and the thermal conductivities are 
\bea
\alpha_L&\equiv&\alpha_{11}(r_h)=\alpha_{22}(r_h)=\f{4\pi \mu(d-2)w_0}{\psi_0k^2L}\left(\f{r_h}{L}\right)^{d-2},\nn
\kappa_L&\equiv& \kappa_{11}(r_h)=\kappa_{22}(r_h)=\left[\f{16\pi^2 T_H  w_0r^{d-1}_h}{L^{d-3}((d-2)^2\mu^2w^2_0+ w_0\psi_0k^2L^2)}\right]\nn
&=&\f{16\pi^2 T_H  }{\psi_0k^2}\left(\f{r_h}{L}\right)^{d-1}\left[1+\f{(d-2)^2w_0\mu^2}{\psi_0k^2L^2}\right]^{-1},\nn
\vartheta_L&=&\vartheta_{11}(r_h)=-\f{4\pi\rho}{w_0\psi k^2}\left(\f{r_h}{L}\right)^{3-d}\left[1+\f{(d-2)^2w_0\mu^2}{\psi_0k^2L^2}\right]^{-1}.
\eea

The temperature  has the following dependence on the chemical potential and the size of the horizon

\be
T_H=\f{dr_h}{4\pi L^2}\left[1-\f{\psi_0k^2L^4}{2dr^2_h}-\f{\mu^2(d-2)^2L^{2}w_0}{2d(d-1)r^{2}_h} \right].
\ee 

The size of the horizon, $r_h$,  can solved in terms of temperature, $T_H$, chemical potential, $\mu$, and the dissipation parameter, $k$,   which then  allows us to  express the result of electrical conductivity as written in eq(\ref{andrade_cond}),    completely in terms of temperature, $T_H$, chemical potential, $\mu$, and the dissipation parameter, $k$. On solving 
\be
\f{r_h}{L}=\mu L\left[\f{2\pi}{d}  \left(\f{T_H}{\mu}\right)+\sqrt{\f{(d-2)^2}{2d(d-1)}\f{w_0}{ L^2}+\f{4\pi^2}{d^2}\left (\f{T_H}{\mu}\right)^2+\f{\psi_0}{2d}\left(\f{k}{\mu}\right)^2}\right],
\ee
where we have taken the higher size of the horizon. This gives the transport coefficients as
\bea
\sigma_L&=&(\mu L)^{d-3}w_0 ~\left(g\bigg( \f{T_H}{\mu}, \f{k}{\mu} \bigg)\right)^{d-3
}\left[1+\f{(d-2)^2w_0\mu^2}{\psi_0k^2L^2}\right],\nn
\alpha_L&=&(\mu L)^{d-1} ~\left(g\bigg( \f{T_H}{\mu}, \f{k}{\mu} \bigg)\right)^{d-2} \left[\f{4\pi(d-2)w_0}{\psi_0k^2 L^2}\right],\nn
\kappa_L&=&(\mu L)^{d-1} ~\left(g\bigg( \f{T_H}{\mu}, \f{k}{\mu} \bigg)\right)^{d-1}\left(\f{16\pi^2 T_H  }{\psi_0k^2}\right)\left[1+\f{(d-2)^2w_0\mu^2}{\psi_0k^2L^2}\right]^{-1},\nn
\vartheta_L&=&-\f{4\pi\rho}{w_0\psi k^2}(\mu L)^{3-d} ~\left(g\bigg( \f{T_H}{\mu}, \f{k}{\mu} \bigg)\right)^{3-d}\left[1+\f{(d-2)^2w_0\mu^2}{\psi_0k^2L^2}\right]^{-1}
\eea
where the function 
\be\label{function_g}
g\bigg( \f{T_H}{\mu}, \f{k}{\mu}  \bigg)=\left[\f{2\pi}{d}  \left(\f{T_H}{\mu}\right)+\sqrt{\f{(d-2)^2}{2d(d-1)} \f{w_0}{L^2}+\f{4\pi^2}{d^2}\left (\f{T_H}{\mu}\right)^2+\f{\psi_0}{2d}\left(\f{k}{\mu}\right)^2}\right].
\ee
 
 For $d=3$, something special happens,  it is possible to separate the incoherent particle-hole pairs from the momentum dissipation due to lattice only for the longitudinal electrical conductivity. 
 
 \paragraph{A holographic relation:} There exists a relation among the transport coefficients. This follows from eq(\ref{transport_zero_B}) 
 \be
 \f{\mu^2}{k^2}\f{\kappa_L\sigma_L}{T_H \alpha^2_L}=\f{W(\phi(r_h))\psi(\phi(r_h)) h^{d-2}(r_h)}{\left(\int^{\infty}_{r_h}\f{\sqrt{U_1(r)}}{\sqrt{U_2(r)}W(\phi(r))h^{\f{d-1}{2}}(r)}\right)^2}
 %=W(\phi(r_h))\psi(\phi(r_h)) h^{d-2}(r_h)\chi^2_e,
 \ee
%where $\chi_e$ is the charge susceptibility. 
For AdS spacetime as written in eq(\ref{ads_spacetime})this gives 
%the charge susceptibility, $\chi_e=\f{L}{(d-2)w_0 h^{\f{d-2}{2}}(r_h)}$. This gives
\be
\f{\mu^2}{k^2}\f{\kappa_L\sigma_L}{T_H \alpha^2_L}=\f{\psi_0 L^2}{(d-2)^2 w_0}={\rm constant}.
\ee

\paragraph{With Magnetic field: } Let us calculate the transport coefficients in the presence of magnetic field for an asymptotically AdS spacetime with  $h(r)=r^2/L^2$. The charge density and the chemical potential are related as given in eq(\ref{charge_density_chemical_potential}). 

The transport coefficients reads with the choice $16\pi G=1$ as 
\bea\label{ads_transport}
\sigma_{11}(r_h)&=&\sigma_{22}(r_h)= \left[\f{k^2w_0\psi_0L^2\left((d-2)^2w_0\mu^2+ \psi_0k^2L^2+\f{B^2L^4w_0}{r^{2}_h}\right)}{ \psi^2_0k^4L^4+\f{B^2L^4w_0}{r^{2}_h}\left((d-2)^2w_0\mu^2+2\psi_0 k^2L^2+\f{B^2L^4w_0}{r^{2}_h}\right)}\right]\left(\f{r_h}{L}\right)^{d-3}\nn
\sigma_{12}(r_h)&=&-\sigma_{21}(r_h)=\left[\f{(d-2)\mu BLw^2_0\left((d-2)^2w_0\mu^2+ 2\psi_0k^2L^2+\f{B^2L^4w_0}{r^{2}_h}\right)}{ \psi^2_0k^4L^4+\f{B^2L^4w_0}{r^{2}_h}\left((d-2)^2w_0\mu^2+2\psi_0 k^2L^2+\f{B^2L^4w_0}{r^{2}_h}\right)}\right]\left(\f{r_h}{L}\right)^{d-4}\nn
\alpha_{11}(r_h)&=&\alpha_{22}(r_h)={\overline\alpha}_{11}(r_h)={\overline\alpha}_{22}(r_h)\nn&=& \left[\f{4\pi \mu (d-2)  w_0\psi_0k^2L^3}{\psi^2_0k^4L^4+\f{B^2L^4w_0}{r^{2}_h}\left((d-2)^2w_0\mu^2+2\psi_0 k^2L^2+\f{B^2L^4w_0}{r^{2}_h}\right) }\right]\left(\f{r_h}{L}\right)^{d-2}\nn
\alpha_{12}(r_h)&=&-\alpha_{21}(r_h)={\overline\alpha}_{12}(r_h)=-{\overline\alpha}_{21}(r_h)\nn&=& 4\pi B w_0L^2\left[\f{\f{B^2L^4w_0}{r^{2}_h}+(d-2)^2w_0\mu^2+\psi_0 k^2L^2}{\psi^2_0k^4L^4+\f{B^2L^4w_0}{r^{2}_h}\left((d-2)^2w_0\mu^2+2\psi_0 k^2L^2+\f{B^2L^4w_0}{r^{2}_h}\right) }\right]\left(\f{r_h}{L}\right)^{d-3},\nn
\kappa_{11}(r_h)&=&\kappa_{22}(r_h)\nn&=&16\pi^2 T_H L^2\left[\f{((d-2)^2w_0\mu^2+\psi_0 k^2L^2)}{\f{B^2}{r^{2}_h} (d-2)^2w^2_0L^4\mu^2  +((d-2)^2w_0\mu^2+ \psi_0k^2L^2)^2}\right]\left(\f{r_h}{L}\right)^{d-1}\nn
\kappa_{12}(r_h)&=&-\kappa_{21}(r_h)\nn&=&-\left[  \f{16\pi^2 B T_H\mu(d-2)w_0 L^3}{\f{B^2}{r^{2}_h} (d-2)^2w^2_0L^4\mu^2  +((d-2)^2w_0\mu^2+ \psi_0k^2L^2)^2}\right]\left(\f{r_h}{L}\right)^{d-2}.
\eea

The temperature in the presence of magnetic reads as
\be
T_H=\f{dr_h}{4\pi L^2}\left[1-\f{B^2w_0L^6 }{4dr^4_h}-\f{k^2\psi_0 L^4}{2dr^2_h}-\f{\mu^2(d-2)^2L^2w_0}{2d(d-1)r^{2}_h} \right].
\ee

In principle, we can solve the size of the horizon as function of the temperature, chemical potential, magnetic field and the dissipation parameter. However, even for the  case $d=3$, the expression of the size of the horizon becomes too messy. So, instead of finding the precise dependence of the transport coefficients on these parameters, we shall plot these coefficients.

Introducing dimensionless variables $(b,~{\tilde\rho},~x_h,~{\tilde k},~t_H~{\tilde \mu})$ as
\be
b\equiv BL\sqrt{w_0},\quad {\tilde\rho}\equiv\f{\rho L}{\sqrt{w_0}},\quad x_h\equiv\f{r_h}{L},\quad {\tilde k}\equiv k L,\quad {\tilde \mu}=\mu \sqrt{w_0}
\ee
leads to 
\bea
t_H&\equiv& T_H L=\f{dx_h}{4\pi }\left[1-\f{b^2 }{4dx^4_h}-\f{{\tilde k}^2\psi_0 }{2dx^2_h}-\f{{\tilde\rho}^2}{2d(d-1)x^{2(d-1)}_h} \right]\nn
{\widetilde\sigma}_{11}&\equiv&\f{\sigma_{11}(x_h)}{w_0}=\f{\sigma_{22}(x_h)}{w_0}= \left[\f{{\tilde k}^2\psi_0\left((d-2)^2{\tilde \mu}^2+ \psi_0{\tilde k}^2+\f{b^2}{x^{2}_h}\right)}{ \psi^2_0{\tilde k}^4+\f{b^2}{x^2_h}\left((d-2)^2{\tilde \mu}^2+2\psi_0 {\tilde k}^2+\f{b^2}{x^{2}_h}\right)}\right]x^{d-3}_h\nn
{\widetilde\sigma}_{12}&\equiv&\f{\sigma_{12}(x_h)}{w_0}=-\f{\sigma_{21}(x_h)}{w_0}=\left[\f{(d-2){\tilde \mu} b\left((d-2)^2{\tilde \mu}^2+2\psi_0 {\tilde k}^2+\f{b^2}{x^{2}_h}\right)}{ \psi^2_0{\tilde k}^4+\f{b^2}{x^2_h}\left((d-2)^2{\tilde \mu}^2+2\psi_0 {\tilde k}^2+\f{b^2}{x^{2}_h}\right)}\right]x^{d-4}_h\nn
{\widetilde\alpha_{11}}&\equiv&\f{\alpha_{11}(x_h)}{\sqrt{w_0}L}=\f{\alpha_{22}(x_h)}{\sqrt{w_0}L}=\f{{\overline\alpha}_{11}(x_h)}{\sqrt{w_0}L}=\f{{\overline\alpha}_{22}(x_h)}{\sqrt{w_0}L}\nn&=& \left[\f{4\pi {\tilde \mu} (d-2)  \psi_0{\tilde k}^2}{ \psi^2_0{\tilde k}^4+\f{b^2}{x^2_h}\left((d-2)^2{\tilde \mu}^2+2\psi_0 {\tilde k}^2+\f{b^2}{x^{2}_h}\right)}\right]x^{d-2}_h\nn
{\widetilde\alpha_{12}}&\equiv&\f{\alpha_{12}(r_h)}{\sqrt{w_0}L}=-\f{\alpha_{21}(r_h)}{\sqrt{w_0}L}=\f{{\overline\alpha}_{12}(r_h)}{\sqrt{w_0}L}=-\f{{\overline\alpha}_{21}(r_h)}{\sqrt{w_0}L}\nn&=& 4\pi b \left[\f{\f{b^2}{x^{2}_h}+(d-2)^2{\tilde \mu}^2+\psi_0 {\tilde k}^2}{ \psi^2_0{\tilde k}^4+\f{b^2}{x^2_h}\left((d-2)^2{\tilde \mu}^2+2\psi_0 {\tilde k}^2+\f{b^2}{x^{2}_h}\right)}\right]x^{d-3}_h,\nn
{\widetilde\kappa}_{11}&\equiv&\f{\kappa_{11}(r_h)}{L}=\f{\kappa_{22}(r_h)}{L}=16\pi^2 t_H \left[\f{((d-2)^2{\tilde \mu}^2+\psi_0 {\tilde k}^2)}{\f{b^2}{x^{2}_h} (d-2)^2{\tilde \mu}^2  +((d-2)^2{\tilde \mu}^2+ \psi_0{\tilde k}^2)^2}\right]x^{d-1}_h\nn
{\widetilde\kappa}_{12}&\equiv&\kappa_{12}(r_h)=-\kappa_{21}(r_h)=-16\pi^2t_H\left[  \f{ b {\tilde \mu}(d-2)}{\f{b^2}{x^{2}_h} (d-2)^2{\tilde \mu}^2  +((d-2)^2{\tilde \mu}^2+ \psi_0{\tilde k}^2)^2}\right]x^{d-2}_h.
\eea

\begin{figure}
	\centering
	\subfigure[]{\includegraphics[width=0.4\textwidth]{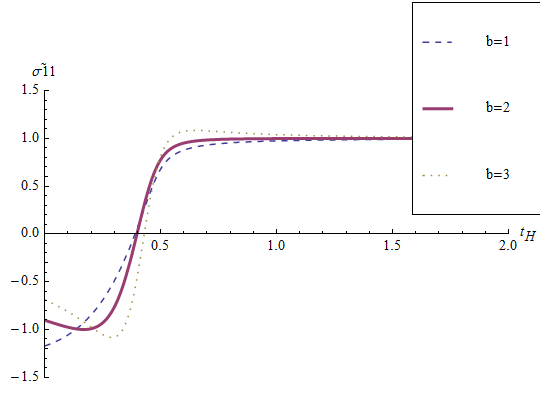}} 
	\subfigure[]{\includegraphics[width=0.4\textwidth]{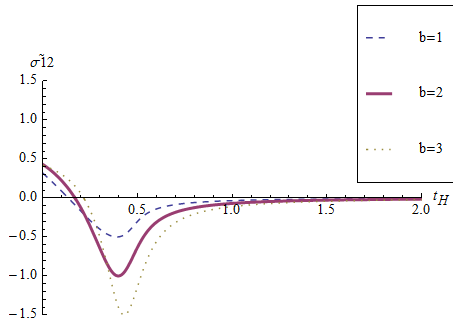}} 
	\subfigure[]{\includegraphics[width=0.4\textwidth]{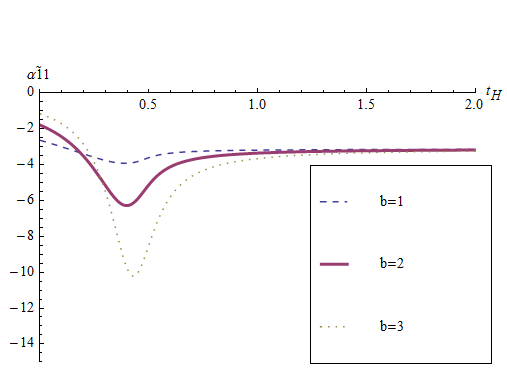}}
	\subfigure[]{\includegraphics[width=0.4\textwidth]{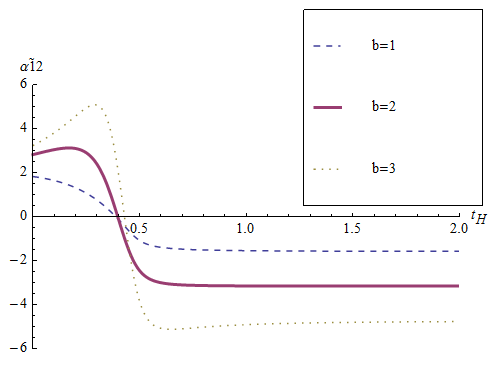}}
	\caption{(a) Longitudinal electrical conductivity  (b) transverse electrical conductivity (c) longitudinal thermo-electrical conductivity (d) transverse thermo-electrical conductivity are plotted versus temperature, $t_H$,  by fixing ${\tilde k}=1,~{\tilde\rho}=2,~\psi_0=-8,~d=3$ for different values of magnetic field, $b$.}
	\label{fig1}
\end{figure}

\begin{figure}
	\centering
	\subfigure[]{\includegraphics[width=0.4\textwidth]{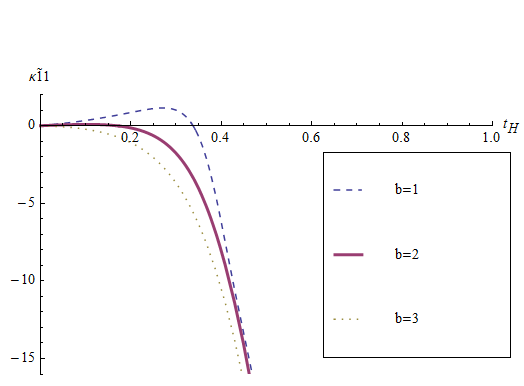}}
	\subfigure[]{\includegraphics[width=0.4\textwidth]{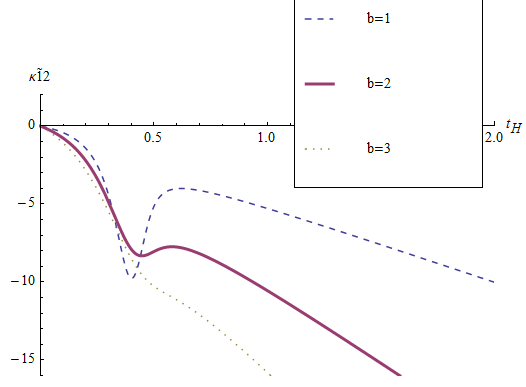}}
	\subfigure[]{\includegraphics[width=0.4\textwidth]{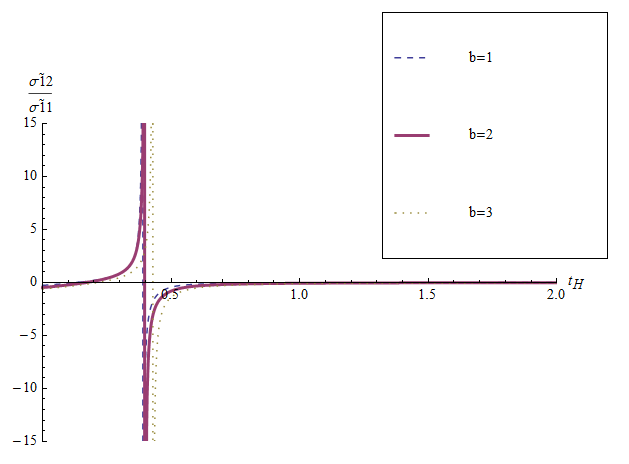}}
	\subfigure[]{\includegraphics[width=0.4\textwidth]{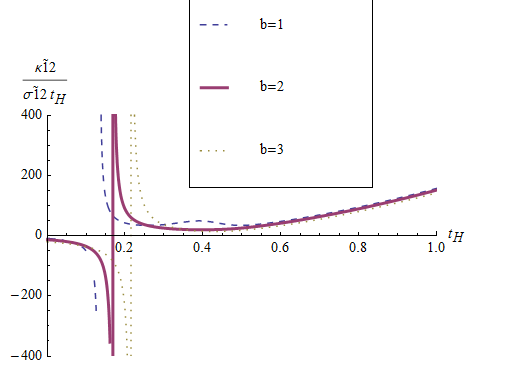}}
	\caption{(a) Longitudinal thermal conductivity  (b) transverse thermal conductivity (c) Hall angle conductivity (d) transverse Lorentz ratio conductivity are plotted versus temperature, $t_H$,  by fixing ${\tilde k}=1,~{\tilde\rho}=2,~\psi_0=-8,~d=3$ for different values of magnetic field, $b$.}
	\label{fig2}
\end{figure}

\begin{figure}
	\centering
	\subfigure[]{\includegraphics[width=0.4\textwidth]{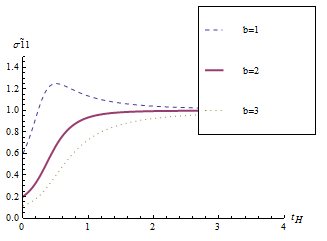}} 
	\subfigure[]{\includegraphics[width=0.4\textwidth]{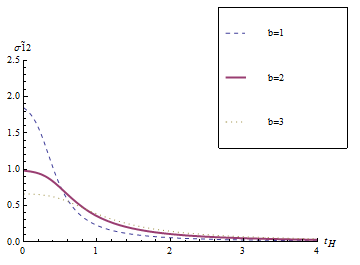}} 
	\subfigure[]{\includegraphics[width=0.4\textwidth]{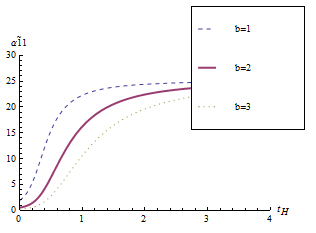}}
	\subfigure[]{\includegraphics[width=0.4\textwidth]{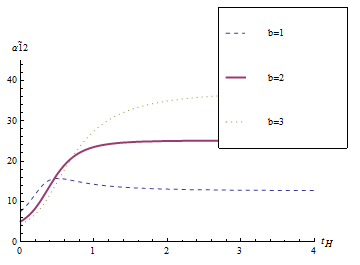}}
	\caption{(a) Longitudinal electrical conductivity  (b) transverse electrical conductivity (c) longitudinal thermo-electrical conductivity (d) transverse thermo-electrical conductivity are plotted versus temperature, $t_H$,  by fixing ${\tilde k}=1,~{\tilde\rho}=2,~\psi_0=1,~d=3$ for different values of magnetic field, $b$.}
	\label{fig3}
\end{figure}

\begin{figure}
	\centering
	\subfigure[]{\includegraphics[width=0.4\textwidth]{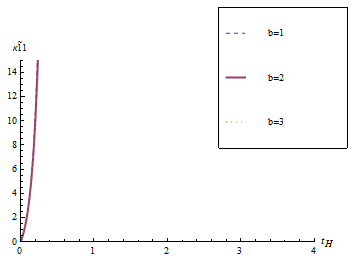}}
	\subfigure[]{\includegraphics[width=0.4\textwidth]{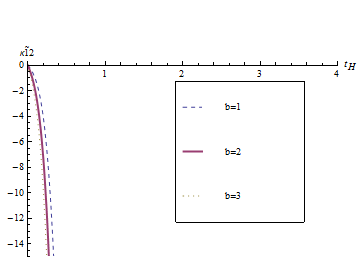}}
	\caption{(a) Longitudinal thermal conductivity  (b) transverse thermal conductivity  are plotted versus temperature, $t_H$,  by fixing ${\tilde k}=1,~{\tilde\rho}=2,~\psi_0=1,~d=3$ for different values of magnetic field, $b$.}
	\label{fig4}
\end{figure}

It follows from the expression of the electrical transport coefficients that for positive values of $\psi_0$ the insulating behavior of AdS spacetime is ruled out. We have plotted the transport coefficients versus temperature, $t_H$, by setting $\psi_0$ to be negative in fig(\ref{fig1}) and fig(\ref{fig2}). Once, we set $\psi_0$ to be negative then it follows that the transport coefficients can become negative as well, whose interpretation is not clear to the author. For completeness, we  have, also,  plotted the transport coefficients versus temperature, $t_H$, by setting $\psi_0$ to be positive in fig(\ref{fig3}) and fig(\ref{fig4}).



\subsection{At IR}

\paragraph{An exact solution without magnetic field:} An exact real valued  solution at IR  follows in the absence of  magnetic field upon  choosing  the potential $V$, the couplings, geometry, gauge potential and the dilaton as

\bea\label{scale_violating}
V(\phi)&=&v_0,\quad \psi(\phi)=\f{8r^2_h}{(d-2)k^2}\equiv \f{\psi_0r^2_h}{k^2},\quad W(\phi)=\f{w_0}{r^2_h}r^{\f{2z(d-2)}{d-1}},\nn\phi(r)&=&\f{\sqrt{2z}\sqrt{2(d-1)-z(d-2)}}{\sqrt{d-1}}~Log~r,\quad A'_t(r)=-\f{4r^2_h(2(d-1)-z(d-2))}{(d-2)\rho~~r^3},\nn
U_1&=&r^{2(\f{d(z-1)-2z+1}{d-1})}\left(1-\left(\f{r_h}{r}\right)^2\right),\quad U_2(r)=r^{2(\f{2(d-1)+z}{d-1})}\left(1-\left(\f{r_h}{r}\right)^2\right),\nn h(r)&=&r^{\f{-2z}{d-1}},\quad w_0=-\f{(d-2)\rho^2}{4(2(d-1)-z(d-2))}
\eea
where $v_0$ is  constant and arbitrary.
The geometry is now described by Lifshitz dynamical exponent $z$ \cite{Kachru:2008yh} and hyperscale violating parameter $\gamma=\f{2(d+z-1)}{d-1}\equiv \f{\theta}{d-1}$ \cite{Huijse:2011ef} and reads as
\be
ds^2
=r^{\f{-2(d+z-1)}{d-1}}\left[-r^{2z}f(r)dt^2+r^2dx^i dx_i+\f{dr^2}{r^2f(r)} \right],\quad f(r)=1-\left(\f{r_h}{r}\right)^2
\ee
The scale violating  parameter, $\gamma$, for the Einstein-Maxwell-dilaton-axion system is not an independent parameter rather  depends on the number of spatial dimensions as well as on the Lifshitz dynamical exponent, $z$, whereas for the Einstein-DBI-dilaton-axion system \cite{Pal:2020gsq}, it is a free parameter. 

In order to have a real valued solution, there is a constraint on the dynamical exponent and it  should be $0\leq z \leq \f{2(d-1)}{d-2}$ for $d\geq 2.$ The Hawking temperature for such a black hole is given by 
\be
T_H=\f{r^z_h}{2\pi}.
\ee

The chemical potential for the scale-violating solution eq(\ref{scale_violating}) reads as
\be
\mu=- \left[\f{4r^2_h(2(d-1)-z(d-2))}{(d-2)\rho}\right]\int^{\infty}_{r_h}\f{dr}{r^3}=-\f{2[2(d-1)-z(d-2)]}{(d-2)\rho}
\ee

Hence, the conductivities are

\bea
\sigma_{11}(r_h)&=&\sigma_{22}(r_h)=h^{\f{d-3}{2}}(r_h)\f{W}{16\pi G}\left[1+\f{\rho^2 h^{2-d}}{W\psi k^2}\right]_{r_h}=\f{w_0}{16\pi G}r^{z-2}_h\left[1+\f{\rho^2}{w_0\psi_0}\right],\nn&=& (2\pi T_H)^{\f{z-2}{z}}\f{w_0}{16\pi G}\left[1+\f{\rho^2}{w_0\psi_0}\right],\nn
\alpha_{11}(r_h)&=&\alpha_{22}(r_h)=\f{1}{16\pi G}\left[\f{4\pi \rho}{\psi_0 r^2_h }\right]=\f{ \rho}{4G\psi_0} (2\pi T_H)^{-\f{2}{z}},\nn
\kappa_{11}(r_h)&=&\kappa_{22}(r_h)=\f{1}{16\pi G}\left[\f{16\pi^2 T_H  h^{\f{d-1}{2}}}{\f{\rho^2 h^{2-d}}{W}+\psi k^2}\right]_{r_h}=\f{1}{2Gr^2_h(\f{\rho^2}{w_0}+\psi_0)}\nn&=&\f{1}{2G\psi_0(1+\f{\rho^2}{w_0\psi_0})} (2\pi T_H)^{-\f{2}{z}}.
\eea 

\paragraph{Interesting relations:} Given the exact result of the longitudinal electrical conductivity, thermo-electrical conductivity and the thermal conductivity, we can calculate the ratio between temperature times the thermal conductivity with the   electrical conductivity and it reads as
\be
\f{T_H \kappa_{11}(r_h)}{ \sigma_{11}(r_h)}=\f{4}{w_0\psi_0}={\rm constant}.
\ee

There exists another interesting relation, which is the ratio between  the thermal conductivity with the   thermo-electrical conductivity and it reads as
\be
\f{ \kappa_{11}(r_h)}{ \alpha_{11}(r_h)}=\f{2}{\rho(1+\f{\rho^2}{w_0\psi_0})}={\rm constant}.
\ee

The entropy density and the specific heat at constant charge density has the following dependence on the dynamical exponent
\be
s=\f{r^{-z}_h}{4G}=\f{1}{8\pi G} T^{-1}_H,\quad C_{\rho}\sim -T^{-1}_H.
\ee
The negative specific heat   suggests this phase at IR is an unstable phase. 
The energy of the system obtained via the first law of thermodynamics  has a logarithmic dependence on temperature
\be
E=-\f{V_{d-1}}{8\pi G} \left[Log(2\pi T_H)+ \left(\f{2(d-1)-z(d-2)}{(d-2)}\right)Log\rho\right].
\ee

\subsection{An exact solution with magnetic field} 

A gravitational solution at finite temperature  that breaks the scaling symmetry is obtained in  \cite{Huijse:2011ef}. The scaling symmetry is broken with the help of a  dilaton field that goes logarithmically. Essentially, it is the back reaction of the dilaton field breaks the scaling symmetry of the geometry. The gravitational solution at finite temperature and  with finite  charge density is obtained with the help of more than one U(1) gauge fields in \cite{Tarrio:2011de, Alishahiha:2012qu}. In this model the magnetic field is included in  \cite{Ge:2016sel}.

In this subsection, we shall obtain a gravitational solution at finite temperature, finite chemical potential in the presence of a constant magnetic field. The interesting point about the solution 
is that it 
will be generated by solving the necessary equations for constant dilaton field and with one U(1) gauge potential.  More importantly, the geometry obtained will break the scaling symmetry.

We can solve eq(\ref{eom}) to find an exact solution  with a magnetic field and it takes the following form
\bea
ds^2&=&r^{-2\gamma}\left[-r^{2 z}f(r)dt^2+r^{2(z-\gamma)} dx^2_i+\f{dr^2}{r^2f(r)}\right],\nn
\phi(r)&=&\phi_0,\quad V(\phi)=-v_0,\quad \psi(\phi)=\psi_0,\quad W(\phi)=w_0,\nn
A'_t(r)&=&\f{\rho}{w_0}r^{-1-(d-2)(z-2\gamma)}, 
\eea
where $\phi_0,~v_0,~\psi_0$ and $w_0$ are constants. The function 
\bea
f(r)&=&\f{v_0}{d(d-1)(z-2\gamma)^2}r^{-2\gamma}\left[1-\left(\f{r_h}{r}\right)^{d(z-2\gamma)}\right]-\f{\psi_0k^2}{2(d-2)(z-2\gamma)^2}r^{-2(z-\gamma)}\left[1-\left(\f{r_h}{r}\right)^{(d-2)(z-2\gamma)}\right]\nn&+&\f{\rho^2}{2w_0(d-1)(d-2)(z-2\gamma)^2}\f{r^{-zd+2\gamma(d-1)}}{r^{(d-2)(z-2\gamma)}_h}\left[1-\left(\f{r_h}{r}\right)^{(d-2)(z-2\gamma)}\right]\nn&-&\f{B^2w_0}{4(d-4)(z-2\gamma)^2}
r^{-2(2z-3\gamma)}\left[1-\left(\f{r_h}{r}\right)^{(d-4)(z-2\gamma)}\right].
\eea
Note, in order to have a non-singular solution, we shall be restricting   $z\neq 2\gamma$. 
   The Hawking temperature for such a case is
\be
T_H=\f{1}{4\pi(z-2\gamma)}\left[\f{v_0r^{z-2\gamma}_h}{(d-1)}-\f{B^2w_0}{4}r^{-3(z-2\gamma)}_h-\f{\rho^2}{2w_0(d-1)}r^{-(2d-3)(z-2\gamma)}_h-\f{\psi_0k^2}{2r^{z-2\gamma}_h}\right].
\ee

The chemical potential 
\be
\mu=\f{\rho}{w_0(d-2)(z-2\gamma)}r^{-(d-2)(z-2\gamma)}_h
\ee

The entropy, charge and the energy of the system is
\bea
S&=&\f{V_{d-1}}{4G}r^{(d-1)(z-2\gamma)}_h,\quad Q= \f{V_{d-1}}{16\pi G}\rho\nn
E&=&\f{V_{d-1}}{16\pi G(z-2\gamma)}\bigg[\f{v_0r^{d(z-2\gamma)}_h}{d}-\f{B^2w_0(d-1)}{4(d-4)}r^{(d-4)(z-2\gamma)}_h\nn&+&\f{\rho^2}{2w_0(d-2)}r^{-(d-2)(z-2\gamma)}_h-\f{\psi_0k^2(d-1)}{2(d-2)}r^{(d-2)(z-2\gamma)}_h\bigg].
\eea

It follows that the free energy, $F=E-T_H S -\mu Q$, reads as
\bea
F&=&-\f{V_{d-1}}{16\pi G(z-2\gamma)}\bigg[\f{v_0r^{d(z-2\gamma)}_h}{d(d-1)}+\f{3B^2w_0}{4(d-4)}r^{(d-4)(z-2\gamma)}_h\nn&+&\f{\rho^2}{2w_0(d-1)(d-2)}r^{-(d-2)(z-2\gamma)}_h+\f{\psi_0k^2}{2(d-2)}r^{(d-2)(z-2\gamma)}_h\bigg]
\eea
From which follows the zero field magnetic susceptibility 
\bea
\chi_B&=&-\p^2_B F=\f{w_0V_{d-1}r^{(d-4)(z-2\gamma)}_h}{32\pi G(d-4)(z-2\gamma)}\times\nn&&\left(\f{2(d-1)v_0r^{2z}_h+r^{4\gamma}_h((d-1)^2\psi_0k^2+(d-7)(d-2)^2w_0 \mu^2(z-2\gamma)^2)}{2v_0r^{2z}_h+r^{4\gamma}_h((d-1)\psi_0k^2-(d-2)^2w_0 \mu^2(z-2\gamma)^2)}\right)_{B=0}.
\eea

\paragraph{The null energy condition:} The null energy condition states that for any null vector $u^M$, the energy momentum tensor obeys $T_{MN}u^M u^N\geq 0$.  Using the Einstein's equation motion, it is easy to show $R_{MN}u^M u^N\geq 0$, where $R_{MN}$ is the Ricci tensor. For the present case, we get $w_0\geq 0$ and $\psi_0\geq 0$.

\paragraph{Transport coefficients:}
The transport coefficients with the choice $16\pi G=1$ are 
\bea\label{transport_b_IR}
\sigma_{11}(r_h)&=&\sigma_{22}(r_h)=\left[\f{\psi_0 k^2r^{(d-3)(z-2\gamma)}_h(\rho^2 r^{-2(d-2)(z-2\gamma)}_h+w_0\psi_0 k^2+w^2_0B^2r^{-2(z-2\gamma)}_h)}{\psi^2_0 k^4+B^2(\rho^2r^{-2(d-1)(z-2\gamma)}_h +2w_0\psi_0 k^2r^{-2(z-2\gamma)}_h+w^2_0B^2r^{-2(z-2\gamma)}_h)}\right]_{r_h}\nn
\sigma_{12}(r_h)&=&-\sigma_{21}(r_h)=\left[\f{\rho Br^{-2(z-2\gamma)}_h(\rho^2 r^{-2(d-2)(z-2\gamma)}_h+2w_0\psi_0 k^2+w^2_0B^2r^{-2(z-2\gamma)}_h)}{\psi^2_0 k^4+B^2(\rho^2r^{-2(d-1)(z-2\gamma)}_h +2w_0\psi_0 k^2r^{-2(z-2\gamma)}_h+w^2_0B^2r^{-2(z-2\gamma)}_h)}\right]_{r_h}\nn
\alpha_{11}(r_h)&=&\alpha_{22}(r_h)={\overline\alpha}_{11}(r_h)={\overline\alpha}_{22}(r_h)\nn&=& \left[\f{4\pi \rho\psi_0 k^2}{\psi^2_0 k^4+B^2(\rho^2r^{-2(d-1)(z-2\gamma)}_h +2w_0\psi_0 k^2r^{-2(z-2\gamma)}_h+w^2_0B^2r^{-2(z-2\gamma)}_h)}\right]_{r_h}\nn
\alpha_{12}(r_h)&=&-\alpha_{21}(r_h)={\overline\alpha}_{12}(r_h)=-{\overline\alpha}_{21}(r_h)\nn&=& \left[ \f{4\pi Br^{(d-3)(z-2\gamma)}_h(\rho^2 r^{-2(d-2)(z-2\gamma)}_h+w_0\psi_0 k^2+w^2_0B^2r^{-2(z-2\gamma)}_h)}{\psi^2_0 k^4+B^2(\rho^2r^{-2(d-1)(z-2\gamma)}_h +2w_0\psi_0 k^2r^{-2(z-2\gamma)}_h+w^2_0B^2r^{-2(z-2\gamma)}_h))}\right]_{r_h}\nn
\kappa_{11}(r_h)&=&\kappa_{22}(r_h)\nn&=&\left[16\pi^2 T_H w_0 r^{(d-1)(z-2\gamma)}_h\f{(\rho^2 r^{-2(d-2)(z-2\gamma)}_h+w_0\psi_0 k^2)}{B^2 w^2_0\rho^2 r^{-2(d-1)(z-2\gamma)}_h +(\rho^2 r^{-2(d-2)(z-2\gamma)}_h+w_0\psi_0 k^2)^2}\right]_{r_h}\nn
\kappa_{12}(r_h)&=&-\kappa_{21}(r_h)=-\left[  \f{16\pi^2 B T_Hw^2_0\rho}{B^2 w^2_0\rho^2 r^{-2(d-1)(z-2\gamma)}_h +(\rho^2 r^{-2(d-2)(z-2\gamma)}_h+w_0\psi_0 k^2)^2}\right]_{r_h}.
\eea
The precise temperature dependence of the transport coefficients are very difficult to find. However, for small magnetic field and charge density, the size of the horizon is related to the temperature as
\be
r^{(z-2\gamma)}_h=\f{1}{2a}\left[T_H\pm\sqrt{T^2_H+4ab} \right]+{\cal O}(\rho^2,~~B^2),
\ee
where $a=\f{v_0}{4\pi(d-1)(z-2\gamma)}$ and $b=\f{\psi_0 k^2}{8\pi(z-2\gamma)}$. In which case, the temperature 
 dependence of the transport coefficients for $d=3$ to leading order in charge density and magnetic fields are  as follows:
\bea
\sigma_{11}(r_h)&=&\sigma_{22}(r_h)=w_0+\f{(\rho^2-B^2)}{\psi_0 k^2}\left(\f{2a}{T_H\pm\sqrt{T^2_H+4ab} }\right)^2+{\cal O}(\rho^4,~~B^4),\nn
\sigma_{12}(r_h)&=&\f{2w_0\rho B}{\psi_0 k^2}\left(\f{2a}{T_H\pm\sqrt{T^2_H+4ab} }\right)^2+{\cal O}(\rho^4,~~B^4),\nn
\alpha_{11}(r_h)&=&\f{4\pi\rho }{\psi_0 k^2}-\f{8\pi\rho B^2}{\psi^2_0 k^4}\left(\f{2a}{T_H\pm\sqrt{T^2_H+4ab} }\right)^2+{\cal O}(\rho^4,~~B^4),\nn
\alpha_{12}(r_h)&=&\f{4\pi w_0 B }{\psi_0 k^2}+\f{4\pi\rho^2 B}{\psi^2_0 k^4}\left(\f{2a}{T_H\pm\sqrt{T^2_H+4ab} }\right)^2+{\cal O}(\rho^4,~~B^4),\nn
\kappa_{11}(r_h)&=&\f{16\pi^2 T_H}{\psi_0 k^2}\left(\f{T_H\pm\sqrt{T^2_H+4ab} }{2a}\right)^2+{\cal O}(\rho^2,~~B^2),\nn
\kappa_{12}(r_h)&=&\f{16\pi^2 T_H\rho B}{\psi^2_0 k^4}+{\cal O}(\rho^2,~~B^2).
\eea

\section{Conclusion}

In this paper, we have revisited the Einstein-Maxwell-dilaton-axion system and have 
studied thermodynamics as well as the transport coefficients associated to an electrically, magnetically charged black hole at finite temperature with planar horizon in arbitrary but even dimensional bulk spacetime. A new black hole solution is obtained at IR which share   properties similar to that of the scale violating solutions like entropy depends on the scale violating parameter as well as on the Lifshitz dynamical exponent through the horizon. One of the distinguishing feature of the new solution is that there is no need to have a  logarithmic profile of the dilaton, instead a constant form of the dilaton will help us to generate the solution. 

We have shown that there exists  dimensionless ratios involving the transport coefficients and reads as
\bea
&&T_H\f{\alpha_{11}(r_h)}{\overline\kappa_{12}(r_h)}\f{\alpha_{12}(r_h)}{\sigma_{11}(r_h)}=1,\nn&& \f{\sigma_{11}(r_h)}{\alpha_{12}(r_h)}\f{\vartheta_{11}(r_h)}{\rho_{11}(r_h)}=-\f{\rho}{16\pi G B}=T_H\f{\vartheta_{11}(r_h)}{\rho_{11}(r_h)}\f{\alpha_{11}(r_h)}{\overline\kappa_{12}(r_h)},\nn &&T_H\f{\alpha_{12}(r_h)}{\sigma_{11}(r_h)}\f{\vartheta_{12}(r_h)}{\kappa_{12}(r_h)}=\f{16\pi G B}{\rho}=\f{\overline\kappa_{12}(r_h)}{\alpha_{11}(r_h)}\f{\vartheta_{12}(r_h)}{\kappa_{12}(r_h)}
\eea
which holds irrespective of the detailed structure of the solution. We have checked that such a relation also holds even for an electrically, magnetically charged black hole at finite temperature with the planar black holes in $d=3$ for  Einstein-DBI-dilaton-axion system. This is discussed in Appendix B.

\section{Appendix: A}
In this section, we shall present the  result  of the on-shell value of the action for the Einstein-Maxwell-dilaton-axion system. This is very much essential to find the thermodynamic potential in grand canonical ensemble. Once, we have the  thermodynamic potential, we can calculate all the thermodynamic quantities. In particular, we are interested to find the magnetization of the system.
It is computed by taking the derivative of the grand potential with respect to the magnetic field for constant temperature,
\be
M=-\f{1}{V_{d-1}\beta}\left(\f{\p I}{\p B} \right)_{\beta},
\ee
where $I,~\beta,~B$ are the grand potential, inverse temperature and magnetic field respectively.  $V_{d-1}=\int d^{d-1}x$ is volume of the spatial directions, $x_i$\rq{}s.

The   Gibbon-Hawking term is
\be
S_{GH}=-\f{1}{\kappa^2}\int d^d x\sqrt{-\gamma} K
\ee
where $K$ is the trace of the extrinsic curvature and $\gamma_{ab}$ is the induced metric.  The extrinsic curvature is defined as $K_{ab}=-\f{1}{2}(\nabla_{a}n_{b}+\nabla_{b}n_{a})$, where $n_{a}$ is the unit vector normal to the boundary.

The sum of the on-shell value of the bulk action and the Gibbon-Hawking term gives
\bea
&&S_{bulk}+S_{GH}=\f{(d-2)V_{d-1}}{2\kappa^2}\left(\beta\f{\sqrt{g_{tt}}}{\sqrt{g_{rr}}}g^{\prime}_{xx}g^{\f{d-3}{2}}_{xx}\right)^{UV}+\f{V_{d-1}}{2\kappa^2}\left(\beta\f{\sqrt{g_{tt}}}{\sqrt{g_{rr}}}g^{\prime}_{xx}g^{\f{d-3}{2}}_{xx}\right)^{IR}\nn &&+\f{V_{d-1}}{2\kappa^2}\left(\beta\f{g^{\prime}_{tt}}{\sqrt{g_{tt}g_{rr}}}g^{\prime}_{xx}g^{\f{d-1}{2}}_{xx}\right)^{UV}-\f{V_{d-1}}{\kappa^2}\beta\int dr \sqrt{g_{tt}g_{rr}}g^{\f{d-3}{2}}_{xx}\left[ \f{\psi}{2}k^2+\f{W}{2g_{xx}}B^2\right].
\eea

The counter terms required   to find a well defined on-shell bulk action on the boundary are

\bea\label{ct_action}
S_{ct1}&=&\f{\alpha_1}{2\kappa^2}\int d^d x \sqrt{-\gamma}=\f{\alpha_1V_{d-1}}{2\kappa^2}
\beta\left(\sqrt{g_{tt}}g^{\f{d-1}{2}}_{xx}\right)^{UV},\nn
S_{ct2}&=&\f{\alpha_2}{\kappa^2}\int d^d x \sqrt{-\gamma} \p_{\mu}\chi_i\p_{\nu}\chi_ig^{\mu\nu}=\f{\alpha_2V_{d-1}}{\kappa^2}(d-1)k^2 \beta\left(\sqrt{g_{tt}}g^{\f{d-3}{2}}_{xx}\right)^{UV}.
\eea
where $\alpha_i$\rq{}s are  constants and the spacetime index is denoted by upper case Latin index $M$, which can $d+1$ values. Here, we shall take $M=(\mu,~r)$, where the Greek index, $\mu$, takes only $d$ values, i.e., defined for the field theory directions.

For the present case, we shall set the metric components as $g_{tt}=U_1(r)=\f{r^2}{L^2}f(r)~~g_{xx}=h(r)=\f{r^2}{L^2}$ and $g_{rr}(r)=\f{1}{U_2(r)}=\f{L^2}{r^2f(r)}$. The couplings are being set as $\psi(\phi(r))=\psi_0$ and $W(\phi(r))=w_0$, where $\psi_0$ and $w_0$ are constants.    The finite term in the sum of  the bulk term, Gibbon-Hawking term and the  counters gives
\bea
S&\equiv& S_{bulk}+S_{GH}+S_{ct1}+S_{ct2},\nn
&=&-\f{V_{2}}{2\kappa^2}\beta\f{(c_1 r_h +B^2 w_0L^6-k^2r^2_h\psi_0L^4)}{ L^4r_h},\quad {\rm for}\quad d=3, ~\alpha_1=-\f{4}{L},~\alpha_2=\f{\psi_0 L}{4},\nn
\eea
where $V_{d-1}=\int d^{d-1}x$ is volume of the spatial directions, $x_i$\rq{}s. 
However, for  $d=5$, we need to add other counter terms
\bea
S_{ct3}&=&\f{\alpha_3}{\kappa^2}\int d^d x \sqrt{-\gamma}F_{\mu\nu}F^{\mu\nu}=\f{\alpha_3V_{d-1}}{\kappa^2}(d-1)B^2 \beta\left(\sqrt{g_{tt}}g^{\f{d-5}{2}}_{xx}\right)^{UV},\nn
S_{ct4}&=&\f{\alpha_4}{\kappa^2}\int d^d x \sqrt{-\gamma}(\p_{\mu}\chi_i\p_{\nu}\chi_ig^{\mu\nu})^2=\f{\alpha_4V_{d-1}}{\kappa^2}(d-1)^2k^4 \beta\left(\sqrt{g_{tt}}g^{\f{d-5}{2}}_{xx}\right)^{UV},
\eea
and for $d=7$, we need some more counter terms
\bea
S_{ct5}&=&\f{\alpha_5}{\kappa^2}\int d^d x \sqrt{-\gamma}(\p_{\mu}\chi_i\p_{\nu}\chi_ig^{\mu\nu})(F_{\rho\sigma}F^{\rho\sigma})=\f{\alpha_5V_{d-1}}{\kappa^2}(d-1)^2B^2k^2 \beta\left(\sqrt{g_{tt}}g^{\f{d-7}{2}}_{xx}\right)^{UV},\nn
S_{ct6}&=&\f{\alpha_6}{\kappa^2}\int d^d x \sqrt{-\gamma}(\p_{\mu}\chi_i\p_{\nu}\chi_ig^{\mu\nu})^3=\f{\alpha_6V_{d-1}}{\kappa^2}(d-1)^3k^6 \beta\left(\sqrt{g_{tt}}g^{\f{d-7}{2}}_{xx}\right)^{UV}.
\eea

The on-shell value of the action for $d=5$ and $d=7$ gives
\bea
&S&\equiv S_{bulk}+S_{GH}+S_{ct1}+S_{ct2}+S_{ct3}+S_{ct4}+S_{ct5}+S_{ct6},\nn
&=&-\f{V_{4}}{2\kappa^2}\beta\f{(3c_1-3B^2 r_h w_0L^6-k^2 r^3_h \psi_0L^4)}{3L^6}\quad {\rm for} ~d=5,~\alpha_1=-\f{8}{L},~\alpha_2=\f{\psi_0L}{12},~\alpha_3=\f{w_0L}{8},\nn&&\alpha_4=\f{\psi^2_0L^3}{1152}\nn
&=&-\f{V_{6}}{2\kappa^2}\beta\f{(15c_1-5B^2 r^3_h w_0L^6-3 k^2 r^5_h\psi_0L^4)}{15}\quad {\rm for} ~d=7,~\alpha_1=-\f{12}{L},~\alpha_2=\f{\psi_0L}{20},~\alpha_3=\f{w_0L}{24},\nn
&&\quad\quad\quad\quad ~\alpha_4=\f{\psi^2_0L^3}{4800},~\alpha_5=\f{w_0\psi_0L^3}{2880},\quad \alpha_6=\f{\psi^3_0L^5}{576000}.
\eea

The quantity, $c_1$, can be easily calculated from the condition, $f(r_h)=0$. 
%This gives 
%\bea
%S&=&\f{V_2}{2\kappa^2}\beta\left(\f{4r^4_h w_0-3B^2 w^2_0+\rho^2+2k^2 r^2_h w_0\psi_0}{4r_h w_0}\right)\quad {\rm for}\quad d=3,\nn
%&=&\f{V_4}{2\kappa^2}\beta\left(\f{24r^8_h w_0+18B^2 r^4_hw^2_0+\rho^2+4k^2 r^6_h w_0\psi_0}{24r^3_h w_0}\right)\quad {\rm for}\quad d=5,\nn
%&=&\f{V_6}{2\kappa^2}\beta\left(\f{60r^{12}_h w_0+15B^2 r^8_hw^2_0+\rho^2+6k^2 r^{10}_h w_0\psi_0}{60r^5_h w_0}\right)\quad {\rm for}\quad d=7,\nn
%\eea
 The grand potential, $I$,  when expressed in terms of the chemical potential reads as
\bea
I&=&-\left(\f{V_2}{4G}\right)\left(\f{r^2_h}{L^2}\right)\left(\f{4r^4_h -3B^2 w_0L^6+r^2_hL^2(\mu^2w_0+2k^2 L^2 \psi_0)}{12r^4_h -B^2 w_0L^6-r^2_hL^2(\mu^2w_0+2k^2 L^2 \psi_0)}\right)\quad {\rm for}\quad d=3,\nn
&=&-\left(\f{V_4}{12G}\right) \left(\f{r^4_h}{L^4}\right)\left(\f{24r^4_h +18B^2 w_0L^6+r^2_hL^2(9\mu^2w_0 +4k^2  \psi_0L^2)}{40r^4_h -2B^2 w_0L^6-r^2_hL^2(9\mu^2w_0 +4k^2 L^2 \psi_0)}\right)\quad {\rm for}\quad d=5,\nn
&=&-\left(\f{V_6}{20G}\right)\left(\f{r^6_h}{L^6}\right) \left(\f{60r^{4}_h +15B^2 w_0L^6+r^2_hL^2(25\mu^2 w_0 +6k^2  \psi_0L^2)}{84r^{4}_h -3B^2 w_0L^6-r^2_hL^2(25\mu^2 w_0 +6k^2 L^2 \psi_0)}\right)\quad {\rm for}\quad d=7,\nn
\eea
where we have used the relation $2\kappa^2=16\pi G.$ Various thermodynamic quantities can be calculated from the free energy. The entropy is defined as 
\bea
S&=&\beta\left( \f{\p I}{\p \beta}\right)_{\mu}-I=\f{V_2 r^2_h}{4G L^2}\quad {\rm for}\quad d=3,\nn
&=&\f{V_4 r^4_h}{4GL^4}\quad {\rm for}\quad d=5,\nn
&=&\f{V_6 r^6_h}{4GL^6}\quad {\rm for}\quad d=7.
\eea

The charge is defined as
\bea
Q&=&-\frac{1}{\beta}\left( \f{\p I}{\p \mu}\right)_{\beta}=\f{V_2 w_0\mu r_h}{16\pi G L^2},\quad {\rm for}\quad d=3,\nn
&=&\f{3V_4 w_0\mu r^3_h}{16\pi GL^4},\quad {\rm for}\quad d=5,\nn
&=&\f{5V_6 w_0\mu r^5_h}{16\pi GL^6},\quad {\rm for}\quad d=7.
\eea

Energy is defined as
\bea
E&=&\left( \f{\p I}{\p \beta}\right)_{\mu}-\f{\mu}{\beta}\left( \f{\p I}{\p \mu}\right)_{\beta}=\f{V_2}{32\pi G r_hL^4}\left(4r^4_h +B^2 w_0L^6+r^2_hL^2(w_0\mu^2-2k^2\psi_0L^2)\right), \quad {\rm for}\quad d=3,\nn
&=&\f{V_4}{96\pi GL^6}\left(24r^5_h-6 B^2 r_h w_0L^6+r^3_hL^2(9w_0\mu^2-4k^2\psi_0L^2)\right)\quad {\rm for}\quad d=5,\nn
&=&\f{V_6}{160\pi GL^8}\left(60r^7_h-5 B^2 r^3_h w_0L^6+r^5_hL^2(25w_0\mu^2-6k^2\psi_0L^2)\right)\quad {\rm for}\quad d=7.
\eea
Magnetic moment is defined as 
\bea
m&=&-\frac{1}{\beta}\left( \f{\p I}{\p B}\right)_{\beta}=-\f{B w_0 V_2 L^2}{16\pi G r_h},\quad {\rm for} \quad d=3,\nn
&=&\f{V_4 B w_0 r_h}{8\pi G},\quad {\rm for} \quad d=5,\nn
&=&\f{V_6 B w_0 r^3_h}{16\pi GL^2},\quad {\rm for} \quad d=7.
\eea

It is easy to notice that the dissipation parameter, $k$, enters,  directly  into the expression of energy.  It also comes into all the thermodynamic quantities via the size of the horizon.
The magnetic susceptibility is defined as the ratio of the magnetic moment per unit volume times the applied magnetic field
\bea
\chi=\f{1}{V}\left(\frac{\p m}{ \p B}\right)_{\beta}&=&-\f{ w_0L^2 }{16\pi G r_h},\quad {\rm for} \quad d=3,\nn
&=&\f{  w_0 r_h}{8\pi G},\quad {\rm for} \quad d=5,\nn
&=&\f{  w_0 r^3_h}{16\pi GL^2},\quad {\rm for} \quad d=7,
\eea
where $V$ is the appropriate volume in the corresponding spacetime. It simply follows that the magnetic susceptibility is negative for $d=3$,  whereas positive  for $d=5$ and $d=7$. Note that the magnetic susceptibility depends on the temperature, magnetic field, chemical potential  and the dissipation parameter, $k$, through the size of the horizon, $r_h$, in a very non-trivial way.

It is easy to see that the thermodynamic quantities obey the following relation
\be
\f{I}{\beta}=E-TS-\mu Q.
\ee

The temperature of the system is
\be
T_H=\f{dr_h}{4\pi L^2}\left[1-\f{B^2 L^6w_0}{4dr^4_h}-\f{k^2L^4\psi_0}{2dr^2_h}-\f{
	\rho^2L^{2d}}{2w_0d(d-1)r^{2(d-1)}_h} \right].
\ee
The condition for the inflection point with respect to the size of the horizon is 
\be
\f{\p T_H}{\p r_h}=0=\f{\p^2 T_H}{\p r^2_h}.
\ee
Restricting to $d=3$, and introducing dimensionless variable $(b,~{\tilde\rho},~x_h,~{\tilde k},~t_H,~t_{crit})$ as
\be
b=BL\sqrt{w_0},\quad {\tilde\rho}=\f{\rho L}{\sqrt{w_0}},\quad x_h=\f{r_h}{L},\quad {\tilde k}=k L,\quad t_H= T_H L,\quad t_{crit}=T_{crit} L
\ee
 gives the condition on the size of the horizon and $\psi_0$ as
\be
x_h=x_{crit}=\f{(b^2+{\tilde\rho}^2)^{\f{1}{4}}}{\sqrt{2}},\quad \psi_0=\psi_{crit}=-\f{6\sqrt{b^2+{\tilde\rho}^2}}{{\tilde k}^2}
\ee
At this value the critical temperature is
\be
t_{crit}=\f{\sqrt{2}}{\pi}~(b^2+{\tilde\rho}^2)^{\f{1}{4}}.
\ee

The temperature, $t_{crit}$, is interpreted to be the temperature for which the turning points appear or disappear in the temperature versus the size of the horizon graph and  $x_{crit}$ represent the size of that black hole.

\begin{figure}[h!]
\centering
   {\includegraphics[ width=8cm,height=6cm]{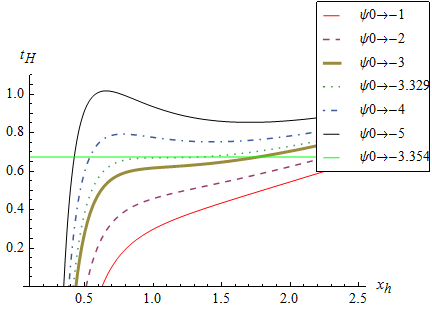} }
  \caption{
   The  figure is   plotted for dimensionless Hawking  temperature, $t_H$, vs the dimensionless size of the horizon, $x_h$, for  $AdS_4$ black hole.  The parameters are set as: $b=2,~d=3,~{\tilde \rho}=1,~{\tilde k}=2$. The $t_{critc}$ is plotted as a  horizontal green line.  }
 \label{fig_1}
\end{figure}

The Hawing-Page phase transition temperature occurs at 
\be
t_{HP}=\f{4b^2-2{\tilde \rho}^2-\psi_0{\tilde k}^2\left(\sqrt{12b^2+\psi^2_0{\tilde k}^4-4{\tilde \rho}^2}-\psi_0{\tilde k}^2 \right)}{\pi\left(\sqrt{12b^2+\psi^2_0{\tilde k}^4-4{\tilde \rho}^2}-\psi_0{\tilde k}^2 \right)^{\f{3}{2}}}
\ee
In order to have a real valued Hawing-Page phase transition temperature, the following condition need to be imposed on the dimensionless magnetic field, charge density and the dissipation parameter
\be
12b^2+\psi^2_0{\tilde k}^4\geq 4{\tilde \rho}^2.
\ee
\section{Appendix B: Transport for DBI system}

The transport coefficients for Einstein-DBI-dilaton-axion system for $d=3$ is calculated  in \cite{Pal:2019bfw, Pal:2020gsq} and restoring the factor of $16\pi G$ reads as
\bea\label{transport_dbi}
&&(16\pi G)\sigma_{11}(r_h)=(16\pi G)\sigma_{22}(r_h)\nn
&=&\left(\f{k^2T_bh Z_2\psi[T_bZ_2(\rho^2+B^2Z^2_1)+k^2\psi\sqrt{\rho^2+Z^2_1(B^2+h^2Z^2_2)}]}{k^4\psi^2(B^2+h^2Z^2_2)+T_bB^2Z_2[T_bZ_2(\rho^2+B^2Z^2_1)+2k^2\psi\sqrt{\rho^2+Z^2_1(B^2+h^2Z^2_2)}]}\right)_{r_h} \nn
&&(16\pi G)\sigma_{12}(r_h)=-(16\pi G)\sigma_{21}(r_h)\nn
&=&\left(\f{B\rho T_b[T^2_bZ^2_2(\rho^2+B^2Z^2_1)+2T_bk^2Z_2\psi\sqrt{\rho^2+Z^2_1(B^2+h^2Z^2_2)}+k^4\psi^2]}{k^4\psi^2(B^2+h^2Z^2_2)+T_bB^2Z_2[T_bZ_2(\rho^2+B^2Z^2_1)+2k^2\psi\sqrt{\rho^2+Z^2_1(B^2+h^2Z^2_2)}]}\right)_{r_h}\nn
&&(16\pi G)T_H \alpha_{11}(r_h)=(16\pi G)T_H \alpha_{22}(r_h)\nn
&=&\left(\f{k^2T_bU_0\rho h^2Z^2_2\psi}{k^4\psi^2(B^2+h^2Z^2_2)+T_bB^2Z_2[T_bZ_2(\rho^2+B^2Z^2_1)+2k^2\psi\sqrt{\rho^2+Z^2_1(B^2+h^2Z^2_2)}]}\right)_{r_h}\nn
&&(16\pi G)T_H \alpha_{12}(r_h)=-(16\pi G)T_H \alpha_{21}(r_h)\nn
&=&\left(\f{B T_bh U_0Z_2[T_bZ_2(\rho^2+B^2Z^2_1)+k^2\psi\sqrt{\rho^2+Z^2_1(B^2+h^2Z^2_2)}]}{k^4\psi^2(B^2+h^2Z^2_2)+T_bB^2Z_2[T_bZ_2(\rho^2+B^2Z^2_1)+2k^2\psi\sqrt{\rho^2+Z^2_1(B^2+h^2Z^2_2)}]}\right)_{r_h}\nn
&&(16\pi G)T_H{\overline\kappa}_{11}(r_h)=(16\pi G)T_H{\overline\kappa}_{22}(r_h)\nn
&=&\left(\f{U^2_0h[k^2\psi(B^2+h^2Z^2_2)+B^2T_bZ_2\sqrt{\rho^2+Z^2_1(B^2+h^2Z^2_2)}]}{k^4\psi^2(B^2+h^2Z^2_2)+T_bB^2Z_2[T_bZ_2(\rho^2+B^2Z^2_1)+2k^2\psi\sqrt{\rho^2+Z^2_1(B^2+h^2Z^2_2)}]}\right)_{r_h} \nn
&&(16\pi G)T_H{\overline\kappa}_{12}(r_h)=-(16\pi G)T_H{\overline\kappa}_{21}(r_h)\nn
&=&\left(\f{\rho BT_bU^2_0h^2Z^2_2}{k^4\psi^2(B^2+h^2Z^2_2)+T_bB^2Z_2[T_bZ_2(\rho^2+B^2Z^2_1)+2k^2\psi\sqrt{\rho^2+Z^2_1(B^2+h^2Z^2_2)}]}\right)_{r_h} \nn
\eea
This gives the longitudinal thermal conductivity as
\bea\label{kappa_11}
&&(16\pi G)\kappa_{11}(r_h)=(16\pi G)\kappa_{22}(r_h)=(16\pi G)\Biggl[{\overline\kappa}_{11}-T_H\f{\left((\alpha^2_{11}-\alpha^2_{12})\sigma_{11}+2\alpha_{11}\alpha_{12}\sigma_{12}\right)}{\sigma^2_{11}+\sigma^2_{12}}\Biggr]_{r_h}\nn
&=&\left(\f{16\pi^2 T_H h\left[T_b \rho^2 Z_2\sqrt{\rho^2+Z^2_1(B^2+h^2 Z^2_2)}+k^2\psi(\rho^2+h^2Z^2_1Z^2_2)\right]}{ [2k^2 T_b\rho^2 Z_2\psi\sqrt{\rho^2+Z^2_1(B^2+h^2 Z^2_2)}+T^2_b\rho^2Z^2_2 (\rho^2+B^2 Z^2_1)+k^4\psi^2(\rho^2+h^2Z^2_1Z^2_2)]}\right)_{r_h}\nn
\eea
The right hand side need to be evaluated at the horizon. The  transverse thermal conductivity takes the following form

\bea\label{kappa_xy}
&&(16\pi G)\kappa_{12}(r_h)=-(16\pi G)\kappa_{21}(r_h)=(16\pi G)\Biggl[{\overline\kappa}_{12}+T_H\f{\left((\alpha^2_{11}-\alpha^2_{12})\sigma_{12}-2\alpha_{11}\alpha_{12}\sigma_{11}\right)}{\sigma^2_{11}+\sigma^2_{12}}\Biggr]_{r_h}\nn&=&-\left(\f{16 \pi^2 T_H B T_b \rho h^2Z^2_1Z^2_2}{ 2k^2 T_b\rho^2 Z_2\psi\sqrt{\rho^2+Z^2_1(B^2+h^2 Z^2_2)}+T^2_b\rho^2Z^2_2 (\rho^2+B^2 Z^2_1)+k^4\psi^2(\rho^2+h^2Z^2_1Z^2_2)}\right)_{r_h}.\nn
\eea

It is easy to show using $U_0=4\pi T_H$ 

\be\label{universal_ratio_dbi}
\f{\alpha_{12}(r_h)}{\sigma_{11}(r_h)}=\left(\f{4\pi B}{\psi k^2}\right)_{r_h},\quad T_H\f{\alpha_{11}(r_h)}{\overline\kappa_{12}(r_h)}=\left(\f{\psi k^2}{4\pi B}\right)_{r_h}
\ee

Upon combining these relations, we get

\be\label{universal_relation_dbi}
T_H\f{\alpha_{11}(r_h)}{\overline\kappa_{12}(r_h)}\f{\alpha_{12}(r_h)}{\sigma_{11}(r_h)}=1.
\ee

The electric charge is defined as $Q=\int d^{d-1}x J^0$, where
\be
J^0=T_b\f{Z_1(\phi) \sqrt{-det\bigg( Z_2(\phi) g_{MN}+\lambda F_{MN}\bigg)}}{32\pi G}\left[\bigg( Z_2(\phi) g+\lambda F\bigg)^{-1~ r0}-\bigg( Z_2(\phi) g+\lambda F\bigg)^{-1~0 r}\right].
\ee

Using the solution to gauge potential as given in \cite{Pal:2020gsq}, gives the electric charge $Q=\f{V_2}{16\pi  G}\rho$. 
It is easy to show that the longitudinal component of the  resistivity matrix is related to the longitudinal component of the Seebeck coefficient as
\bea
\rho_{11}(r_h)&\equiv&\f{\sigma_{11}(r_h)}{\sigma^2_{11}(r_h)+\sigma^2_{12}(r_h)}=-(16\pi G)\left[\f{\psi k^2}{4\pi\rho}\vartheta_{11}(r)\right]_{r_h},\quad {\rm where}\nn
\vartheta_{11}(r_h)&=&-\f{\sigma_{11}(r_h)\alpha_{11}(r_h)+\sigma_{12}(r_h)\alpha_{12}(r_h)}{\sigma^2_{11}(r_h)+\sigma^2_{12}(r_h)}
\eea
This gives the  relation
\be
\f{\sigma_{11}(r_h)}{\alpha_{12}(r_h)}\f{\vartheta_{11}(r_h)}{\rho_{11}(r_h)}=-\f{1}{V_2}\left(\f{Q}{B}\right)=T_H\f{\vartheta_{11}(r_h)}{\rho_{11}(r_h)}\f{\alpha_{11}(r_h)}{\overline\kappa_{12}(r_h)}.
\ee

\end{document}